\documentclass[desactivate]{aa}  

\DeclareRobustCommand{\VAN}[3]{#2}
\let\VANthebibliography\thebibliography
\def\thebibliography{\DeclareRobustCommand{\VAN}[3]{##3}\VANthebibliography}

\usepackage[colorlinks=true,
    linkcolor=blue, citecolor=blue, filecolor=blue, urlcolor=blue]{hyperref}
\usepackage{caption}
\usepackage{subcaption}
\usepackage{graphicx}	
\usepackage{amsmath}	
\usepackage{amssymb}	
\usepackage{array}
\usepackage{multirow}
\usepackage{bigdelim}
\usepackage[varg]{txfonts}
\usepackage{xcolor}
\usepackage{placeins}

\begin{document}

\title{A probabilistic model to estimate number densities from column densities in molecular clouds}
\titlerunning{Probablistic model for GMC densities}
\authorrunning{B.A.L. Gaches \& M. Grudi\'c}

\author{Brandt A. L. Gaches
        \inst{1}\thanks{E-mail: brandt.gaches@chalmers.se}
        \and
        Michael Y. Grudi\'c\inst{2}
        }
\institute{
Department of Space, Earth and Environment, Chalmers University of Technology, Gothenburg SE-412 96, Sweden
\and
Carnegie Observatories, 813 Santa Barbara St, Pasadena, CA 91101, USA
}

\date{Accepted XXX. Received YYY; in original form ZZZ}

\abstract
{Constraining the physical and chemical evolution of molecular clouds is essential to our understanding of star formation. These investigations often necessitate knowledge of some local representative number density of the gas along the line of sight. However, constraining the number density is a difficult endeavor. Robust constraints on the number density often require line observations of specific molecules along with radiation transfer modeling, which provides densities traced by that specific molecule. Column density maps of molecular clouds are more readily available, with many high-fidelity maps calculated from dust emission and extinction, in particular from surveys conduction with the Herschel Space Observatory. We introduce a new probabilistic model which is based on the assumption that the total hydrogen nuclei column density along a line of sight can be decomposed into a turbulent component and a gravitationally dominated component. Therefore, for each pixel in a column density map, the line of sight was decomposed into characteristic diffuse (dubbed ``turbulent'') and dense (dubbed ``gravitational'') gas number densities from column density maps. The method thus exploits a physical model of turbulence to decouple the random turbulent column from gas in dense bound structures empirically using the observed column density maps. We find the model produces reasonable turbulent and gravitational densities in the Taurus L1495/B213 and Polaris Flare clouds. The model can also be used to infer an effective attenuating column density into the cloud, which is useful for astrochemical models of the clouds. We conclude by demonstrating an application of this method by predicting the emission of the [C II] 1900 GHz, [C I] 492 GHz, and CO (J = 1-0) 115 GHz lines across the Taurus L1495/B213 region at the native resolution of the column density map utilizing a grid of photodissociation-region models.}


\maketitle

\section{Introduction}
In the local universe, giant molecular clouds (GMCs) are the primary sites of star formation. GMCs make up the coldest phase of the interstellar medium (ISM), consisting of dense ($\gtrapprox 100$ cm$^{-3}$) and cold ($\approx 20$ K) gas, in which the hydrogen exists primarily in molecular form (H$_2$) \citep{Draine2011, Kennicutt2012}. Understanding the star formation process requires a complete picture of GMCs, in particular the local properties of the gas within the cloud. 

One of the more crucial parameters for understanding the properties of GMCs is the density, $\rho$, as a mass density (g cm$^{-3}$), or $n_H = \rho/(\mu m_H)$ as a hydrogen nuclei number density (cm$^{-3}$), where $\mu$ is the reduced molecular mass, within the cloud. This is an intrinsically difficult value to quantify since observations are typically constrained to line-of-sight-integrated quantities. Therefore, while we can get robust constraints on, for example, line-of-sight-integrated column densities (cm$^{-2}$) or fluxes, constraining volumetric properties is difficult. The gas density throughout a molecular cloud is important since from it one can estimate a wide range of important physical values to understand the dynamics and chemistry. Several crucial processes necessitate constraints on the number density: the free-fall time, which varies as $t_{\rm ff} \propto \rho^{-1/2}$, the virial parameter, through the mass of a structure, $\alpha \propto M^{-1}$, and the Jean's length, $\lambda_J \propto \rho^{-1/2}$, all of which are important to understand the balance between thermal and nonthermal motions and gravity. Analytic and empirical star formation theories inherently require understanding the density through either the free-fall time, virial parameter, or Jean's length \citep[e.g.,][]{McKee2003, Krumholz2005, Hennebelle2008, Padoan2012}. Further, estimates of the magnetic field, in particular using the Davis-Chandrasekhar-Fermi (DCF) method \citep{Davis1951, Chandrasekhar1953} and related methods \citep[see e.g.,][]{Skalidis2021} require an estimation of the density. The plane-of-sky magnetic field determined by DCF is sensitive to density as $B \propto \sqrt{4\pi\rho}$. Finally, chemical models of GMCs require the local density to properly estimate the reaction rates, along with estimations of the effective attenuating column density to accurately model the photochemistry due to external ultraviolet radiation \citep{Tielens2013, Bovino2024}.

There are a variety of methods that have been utilized to estimate the local number density. A particularly robust method is to use a chemical tracer, in particular, the various carbon chains such as C$_2$ and HC$_3$N \citep[e.g.,][]{Federman1994, Sonnentrucker2007, Li2012, Federman2021, Fan2024, Taniguchi2024}. For these different tracers, measurements of different lines can lead to accurate estimations of the local number density. However, these methods produce a biased density - they only estimate the density in regions where the molecule is abundant. Using column density maps, such as the high-resolution maps derived from {\sc Herschel} observations, number densities can be derived assuming uniform density structures and specific geometries, such as spherical or cylindrical masses \citep[see e.g.,][]{Kainulainen2014, Andre2014, Hasenberger2020}. The use of {\sc Gaia} data has enabled the production of three-dimensional dust maps \citep{Leike2019, Leike2020, Zucker2021, Dharmawardena2022, RezaeiKh2022, Edenhofer2024} through which a gas density can be derived following an assumed dust-to-gas mass ratio. Finally, recent innovations in machine learning have enabled an estimation of number density, such as the use of a denoising diffusion probabilistic model (DDPM) using simulated molecular clouds as the training set \citep{Xu2023}, which, promisingly, will allow for number densities to be estimated as training sets grow. 

All of the above methods produce a singular characteristic value for the number density per pixel. For instance, the DDPM method of \citet{Xu2023} estimates a number density based on training data of line-of-sight mass-weighted average densities, while geometric methods such as those from \citet{Kainulainen2014} and \citet{Hasenberger2020} predict geometry-specific volume-averaged densities. While these methods infer some characteristic number density, they are often either agnostic to the underlying physics of GMCs yet depend on geometry assumptions, or they amount to black boxes in which density estimates are provided with a diminished ability to understand underlying errors or biases.

In this paper, we present a new method to estimate characteristic number densities in GMCs. Our new method decomposes the line of sight into a component dominated by turbulence and another dominated by gravity, thus producing two characteristic densities along each pixel. The method can be generalized to produce distributions of these components to enable statistical sampling for forward modeling of emission components. The strength of the method is that the gravitational gas component along each line of sight can be decoupled from the turbulent component and as such models to estimate emission can become more robust by including both the envelope of the turbulent gas and denser gravitationally dominated gas. Further, the method can be utilized to also compute an approximate effective column density, which describes the effective column density that is attenuating an isotropic interstellar radiation field. In Section \ref{sec:methods} we present the underlying physical model and the methodology to estimate number densities from observed column densities. In Section \ref{sec:results} we show a benchmark against a simulation molecular cloud from the {\sc Starforge} suite and estimations for the Taurus and Polaris Flare clouds. Finally, in Section \ref{sec:discussion} we discuss the initial results of the model, along with a discussion of the caveats and conclusions.

\section{Methods}\label{sec:methods}
\subsection{Brief diversion into the N-PDF and n-PDF}
The model proposed here is built upon utilizing known physics encoded in the probability distribution function (PDF) of the line-of-sight integrated column density, denoted N-PDF, or the analog using the number density, denoted n-PDF. Both of these have similar features, broadly consisting of a log-normal component at low density (column density) with a power-law component toward high density (column density) which becomes more dominant over time \citep{Klessen2000,Ballesteros-Paredes2012, Brunt2015,Myers2015, Chen2018, Schneider2022}. The nature of the observed log-normal component has been debated: while theoretical models of turbulence produce such log-normal components \citep[][]{Federrath2008, Burkhart2012}, its location also coincides with the where biases can occur due to noise, background and foreground gas, and statistical completeness \citep{Ossenkopf-Okada2016, Alves2017}. Here, we assume that the log-normal is due to turbulence, but note that our underlying model, as presented, is agnostic to the underlying shape of the N-PDF. 

The properties of the n- and N-PDFs have been well characterized theoretically in previous studies. The width of the log-normal component has been found to depend on the Mach number and nature of the turbulence \citep{Burkhart2012, Federrath2008}. The power-law component is due to self-gravitating gas, with the transition point between the self-gravitating and turbulent components and the slope of the power-law depending on the star formation threshold of the cloud and the star formation rate \citep{Girichidis2014,  Burkhart2018, Chen2018, Burkhart2019,Appel2023}. While these previous studies have decomposed the N-PDF (or n-PDF) into a log-normal turbulent and power-law gravitational component to separate out regions of the cloud into turbulence or gravity-dominated regions, the method presented here uses the information encoded in the N-PDF to decompose each pixel along the line of sight. In this way, it treats the turbulent component of the N-PDF as a bulk probability function to sample the column density, biased by the observed column density at each pixel. Thus, we note that we do not aim to reproduce or explain the N-PDF but to exploit the known physics encoded in it for our line-of-sight decomposition.

\subsection{Physical model}
\begin{figure}[htb!]
    \centering
    \includegraphics[width=0.75\columnwidth]{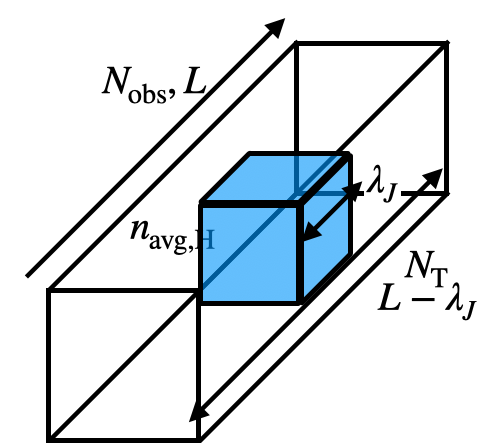}
    \caption{\label{fig:schematic} Schematic for constraining the gas density along the line of sight from an observed column density.}
\end{figure}

We built a simple analytic probabilistic model to relate an observed column density to charactersitic number densities. The general schematic of the model is shown in Figure \ref{fig:schematic}. Along a given line of sight, the column density through the cloud is broken into two components: 1) the turbulence-dominated component along the line of sight, denoted as the ``turbulent column density'' and 2) a gravitationally dominated component, denoted as the ``gravitational column density'' (annotated as the blue cube in Figure \ref{fig:schematic}). 

We assumed that depending on the local density, two regimes dominate the dynamics depending on if the density is less than or greater than some reference transition density, $n_{\rm tr}$: 
\begin{itemize}
    \item $n_H < n_{\rm tr}$, the gas is dominated purely by turbulent motions 
    \item $n_H > n_{\rm tr}$, the gas is dominated by gravitational collapse within a turbulent domain.
\end{itemize}
The transition density here is chosen to be the critical density at which the Jean's length is equal to the sonic scale, $\lambda_J \approx \lambda_s$ \citep{Krumholz2005,Burkhart2019}, where we used the parameterization from \citet{Burkhart2019}, given by
\begin{equation}
    n_{\rm tr} = \frac{\pi c_s^2 M_s^4}{\mu m_H G L^2},
\end{equation}
where $c_s = \sqrt{k_bT/\mu m_H}$ is the thermal sound speed, $M_s$ is the Sonic Mach number of the turbulence, $T$ is the gas temperature, $\mu = 2.33$ is the reduced mass, and $L$ is the size of the cloud.

We described the turbulence component of the line of sight integrated column density as a log-normal, with a width of $\sigma_\zeta$, where we have defined $\zeta = \ln N/N_0$, where $N_0$ is the mean of the log-normal component \citep{Vazquez-Semadeni1994,Padoan1997}. The turbulent column density is then
\begin{equation}\label{eq:NT}
    N_T(n_H) = N_0 \exp \left [\mathcal{N}(0, \sigma_\zeta) \right ],
\end{equation}
where $\mathcal{N}$ is the normal distribution. The statistical distribution represents a stochastic contribution to the line-of-sight column density due to integrating through turbulent gas motions. 

Since one of the fundamental parameters of the model is the Mach number, we can utilize the known properties of the N-PDF caused by turbulence to constrain it. The Gaussian width of the turbulent column density was parameterized as
\begin{equation}\label{eq:sigmaZeta}
    \sigma_{\zeta}^2 \approx A\times \ln \left ( 1 + b^2 M_s^2\right ),
\end{equation}
where $A$ and $b$ are factors describing the turbulence physics \citep{Vazquez-Semadeni1994, Padoan1997, Burkhart2012}. We use $A = 0.16$ and $b = 0.5$ corresponding to the fit from \citet{Burkhart2012} for Taurus as the assumed parameters for the results presented below for the Taurus L1495/B213 and Polaris Flare regions. In the above expression, $A$ is fit and provides the width of the turbulent velocity dispersion with respect to the Mach number and $b$ is sensitive to the underlying mode of turbulence, with $b = 1/3$ giving pure solenoidal and $b = 1$ pure compressive turbulence.

In the high-density regime, the region becomes dominated by gravitational collapse, with some random fluctuations in the density still evident along the line of sight due to turbulence in the surrounding envelope. The length scale of relevance here is the Jean's length, and so the gravitational column density is approximately
\begin{equation}
    N_J(n_H) \approx \lambda_J n_H = \left ( \frac{\pi}{G\mu m_H} \right )^{1/2} c_s n_H^{1/2}.
\end{equation}

These two regimes were combined using a transition switch from the turbulence to gravity-dominated regimes. The final probabilistic model is
\begin{equation}
    N_{\rm obs} = N_T + S(n_H/n_{\rm tr}) \times N_J. 
\end{equation}
The function $S(n_H/n_{\rm tr})$ is a switch to determine when the gas at some density, $n_H$, is dominated by gravitational or turbulent motions. It is worth noting that it is assumed there is always a turbulent envelope, so the switch only determines if, along the line of sight, there is a component that can collapse gravitationally. There are different possible (sigmoid) switches, but one with a particularly favorable behavior is $x/(1+x)$ where $x = n_H/n_{\rm tr}$ since it enables the function to be analytically inverted. The final equation is then
\begin{equation}\label{eq:main}
    N_{\rm obs} = N_0  \exp \left [\mathcal{N}(0, \sigma_\zeta) \right ] + \left ( \frac{n_H/n_{\rm tr}}{1 + n_H/n_{\rm tr}}\right ) \left ( \frac{\pi}{G\mu m_H} \right )^{1/2} c_s n_H^{1/2}.
\end{equation}
Equation \ref{eq:main} has several free parameters: $T$, $M_s$, $N_0$, $b$, $A$, and $L$. However, many of these can be constrained by observations of the cloud complex on the whole, or informed by numerical experiments of turbulent molecular clouds.

\begin{figure}
    \centering
    \includegraphics[width=\columnwidth]{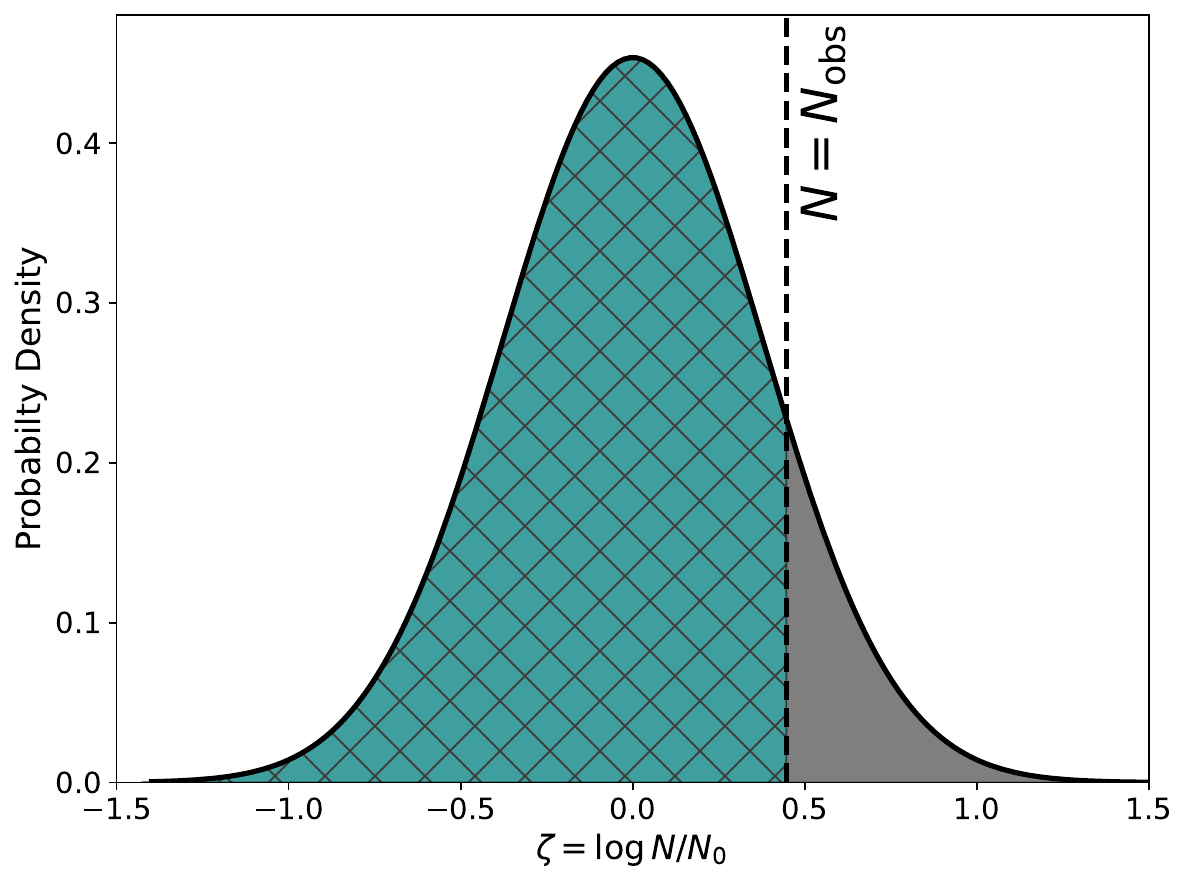}
    \caption{\label{fig:truncNorm}Comparison of the underlying log-normal distribution of the turbulent column density (black) and the sampled truncated log-normal distribution (cyan) for a given observed column density.}
\end{figure}

Starting from Equation \ref{eq:main}, we defined $\Delta N \equiv N_{\rm obs} - N_T$. For a switch function, $S(x) = x/(1+x)$, $\Delta N \propto n_H^{1/2}$, and is purely deterministic with respect to $\Delta N$ (no random sampling for $n_H$ given an $\Delta N$). Observationally, the distribution of $N_T$ along a particular line of sight is biased: it is the log-normal component of the N-PDF constrained by the knowledge that the turbulent column density must be less than or equal to the observed column density. Using $\zeta_{\rm obs} = \ln N_{\rm obs}/N_0$, we utilized a truncated Normal distribution function,
\begin{equation}\label{eq:NTtr}
    \mathcal{N}_{tr}(0, \sigma_\zeta, \zeta_{obs}) = \left [ 1 - H(\zeta - \zeta_{\rm obs}) \right ] \mathcal{N}(0, \sigma_\zeta),
\end{equation} 
where $H(x)$ is the Heaviside step function. This function is sampled using the {\sc Scipy} {\it stats} package truncated normal distribution. Figure \ref{fig:truncNorm} highlights the difference between the underlying distribution for the turbulent column density for the entire cloud and the biased (truncated) turbulent column density that is sampled for a pixel with prior knowledge of the observed column density, $N_{\rm obs}$. From a sampled $N_T$, we compute $\Delta N$ and the corresponding $n_H$, where we denote the number density derived in this way as $n_J$ and denote it as the ``gravitational'' density.

Once $n_J$ is known, it can be used with the macroscopic cloud turbulence properties to estimate a probable effective column density, $N_{\rm eff}$, which is useful for astrochemical models, under the assumption of an isotropic external radiation field. The effective column density is the effective attenuating column density that extincts the external radiation field from the cloud boundary to particular point in the cloud \citep[][]{Glover2010}, in this case, the location of the dense component represented by the blue box in Figure \ref{fig:schematic}. Therefore, this physical setup is similar to the one posed above, with a difference in geometry: from a point in the cloud out, versus the observed column density integrating through the cloud. For a gas parcel of density $n_J$, the effective column will be the addition of the local gravitational column density ($N_J$) and a turbulent column density contribution from the point where the local high-density region is (blue box in Figure \ref{fig:schematic}) out to the cloud boundary. The effective column density is most sensitive to the smallest column density out of the cloud, since the flux from the boundary scales as $\propto e^{-N_{\rm eff}}$, and therefore the shortest gas column density out of the cloud dominates assuming isotropic irradiation. The turbulent component is described in this case as
\begin{equation}\label{eq:NTeff}
    N_{T, {\rm eff}} = \frac{\mathcal{R}}{2} \times N_0 \exp \left [ \mathcal{N}_{tr}(0, \sigma_\zeta, \zeta_{obs}) \right ],
\end{equation}
where the factor of $1/2$ comes from the average minimum turbulent column density being half the total turbulent column density, assuming isotropic turbulence,  $\mathcal{R}$ is a uniform random number between $(0, 1)$ describing the fraction of the spatial depth into the cloud that the densest component is at. The total effective column density is then
\begin{equation}
    N_{\rm eff} = N_J + N_{T, {\rm eff}}.
\end{equation}

\subsection{Applying the method to data}
From a given column density map, we took the N-PDF of the entire region and determine $N_0$ to compute the $\zeta$ distribution. The $\zeta$ distribution is fit with a Gaussian to determine its width which then constrains the Mach number. In our presented results, we assumed the cloud is isothermal, although in principle gas temperatures along the line of sight can also be used. Table \ref{tab:parameters} shows the primary input parameters into the model and their assumed values for the results presented here for observed molecular clouds.

For each pixel in the map, we drew $N_{\rm samp} = 10^3$ samples of $N_T$ using the constraint of the observed column density. In practice, the truncated log-normal in Equation \ref{eq:NTtr} is sampled $N_{\rm samp}$ times for each pixel, using that pixel's observed total column density, to produce a statistical distribution of $N_T$. The number of samples was determined through trial and error balancing of ensuring robust results, memory, and CPU usage. These samples by construction always meet the criterion $\Delta N \geq 0$. The randomly drawn $N_T$ provides $\Delta N$, which is then used to compute $n_J$, and the Jean's length, $\lambda_J$. We defined an effective turbulent gas density $n_T \equiv N_T/(L - \bar{\lambda}_J)$ and an effective gravitational column density, $N_J \equiv \bar{\lambda}_J n_J$, where the effective Jean's length scale is $\bar{\lambda}_J = S(x)\lambda_J(T)$. For each pixel, this produced correlated distributions of the column densities and number densities: ($N_T$, $N_J$), or ($n_T$, $n_J$). 

We present three different choices for utilizing this method. For the first two choices, from the distribution of $N_T$, we utilized the most likely (Method I) or the maximum value (Method II) in the sample. Similarly, we used the mean of $N_{T, {\rm eff}}$ for Method I and the maximum of $N_{T, {\rm eff}}$ for Method II when computing the effective column density. These approaches can be readily used over an entire map to get estimated values across a cloud for the turbulent and gravitational densities and can be computed at the same time. These methods each have their benefits. Method I produces a more diffuse envelope dominated by lower-density gas, but potentially overestimates the gravitational density toward higher column density since the maximum $n_T$ using this method will be approximately $N_0/L$, thus biasing toward higher $\Delta N$. Method II produces more turbulent gas and therefore may overestimate the amount of gas dominated by turbulence in the lower column density region since it naturally assumes that effectively all of the line-of-sight column density is integrated through a purely turbulent medium. For Method II, at higher column densities there is a slight decrease in $n_J$ due to the tail of the log-normal being probed more often, biasing toward lower $\Delta N$. Since both can be computed at the same time, the user can use both to get both lower- and upper-limits of predicted densities and emissions. Finally, for Method III, for given subregions of the cloud, we stored $N_{\rm samp} = 10^4$ samples of $N_T$ to construct fully bivariate distributions of ($n_T$, $n_J$) (chosen by trial and error similar to above, with the greatest constraint being memory usage of this method). For large maps with many pixels, this method can have significant memory overhead but enables a detailed investigation into the distributions. Further, saving the full distributions enables later returning to perform additional computations of other physical parameters using the resulting number and column densities, such as modeling the continuum and line emission.

\begin{table*}
    \caption{Important physical parameters}              
    \label{tab:parameters}     
    \centering                                      
    \begin{tabular}{c|c|c|c|c|c}          
    \hline\hline
    Parameter & Description & Impact variable & Taurus L1495/B213 & Polaris Flare & Starforge \\    
    \hline 
        $L$ (pc) & Cloud Depth & $n_{\rm tr}$, $n_T$ & 1.0$^{1,2}$ & 2.5$^{1,2}$ & 100 (box)\\
        $T$ (K) & Gas Temperature & $c_s$, $n_{\rm tr}$, $\lambda_J$ & 15 & 15 & 15 \\
        $N_0$ (cm$^{-2}$) & Column density peak & $N_T$ & $1.28\times10^{21}$ & $4.35\times10^{21}$ & $2.03\times10^{21}$ \\
        $A$ & Turbulence Model Fit & $\mathcal{M}_s$, $n_{\rm tr}$ & 0.16$^3$ & 0.16$^3$ & no fit \\
        $b$ & Turbulence driving mode & $\mathcal{M}_s$, $n_{\rm tr}$ & 0.5$^3$ & 0.5$^3$ & no fit \\
    \hline 
    \end{tabular}
    \tablebib{(1)~\citet{Qian2015}; (2) \citet{Zucker2021}; (3) \citet{Burkhart2012}}
\end{table*}

\section{Results}\label{sec:results}
We first examine our model predictions using simulation data. Then, we present the predictions for two molecular clouds, representing different regimes of molecular clouds and star formation, the Polaris Flare and the Taurus L1495/B213 region.

\subsection{Comparison against simulations}\label{sec:starforge}
We first compare the results of our methods with a simulation from the {\sc Starforge} suite \citep{Grudic2022}. The simulation uses the {\sc M2e4\_R10\_mu1.3} model from the full physics suite from \citep{Guszejnov2022} including magnetic fields, cooling and chemistry processes, and prescriptions for sink particles, stellar evolution and feedback \citep{Grudic2021}. The simulation solves the equations of ideal magneto-hydrodynamics using the Meshless Finite Mass solver in the {\sc Gizmo} code \citep{Hopkins2015, Hopkins2016}. The initial conditions are a uniform spherical cloud with a total mass of $2\times10^4$ M$_{\odot}$ contained within a 10 pc sphere, placed in the center of a periodic box of size 100 pc. The domain is filled with diffuse gas with a density 1000 times lower than the cloud with a total mass of $5000$ M$_{\odot}$. We refer the reader to \citet{Grudic2022} for further simulation details. The cloud is initialized with a turbulent velocity field. We examine a snapshot when the cloud is 4.9 Myr old when there is ongoing star formation but before the cloud disperses. We compare the turbulent and gravitational densities derived from the simulation column density to the volume- and mass-weighted densities along the line-of-sight to gain intuition on the method's performance and outputs. We perform the analysis using the line-of-sight integrated column density map as if the simulation were an observed cloud. Further, since we know the prescribed initial conditions of the driven turbulence, we do not use Equation \ref{eq:sigmaZeta} and instead use $\mathcal{M}_s = 10$. We must still measure the width of the Gaussian component of the N-PDF, and we find that $\sigma_\zeta \approx 0.72$. The resulting transition number density is $n_{\rm tr} = 1870$ cm$^{-3}$.

Simulations often report two different line-of-sight densities: a volume-weighted average number density and a mass-weighted average number density. These values are useful for simulators to describe their density distributions statistically but are not directly tied to physical meaning. However, since these have been standard, we compare our turbulence and gravitational number densities to these to find how they relate and to gain some more statistical insight into what these components are tracing. The volume-weighted average number density is
\begin{equation}
    n_V = \frac{\sum_i^{N_{\rm cells}} \Delta V_i n_{H,i}}{\sum_i^{N_{\rm cells}} \Delta V_i}.
\end{equation}
For a uniform grid cube ($\Delta x_i = $ constant, therefore $\Delta V_i =$ constant) this becomes
\begin{equation}
    n_V = \frac{1}{N_{\rm cells}} \sum_i^{N_{\rm cells}} n_{H,i},
\end{equation}
where $N_{\rm cells}$ is the number of cells along the line of sight, $\Delta V_i$ is the differential volume element of the cell and $n_{H,i}$ is the hydrogen nuclei number density of the $i$th pixel along the line of sight. The mass-weighted average number density follows from
\begin{equation}
    n_M = \frac{\sum_i^{N_{\rm cells}} m_i n_{H,i}}{\sum_i^{N_{\rm cells}} m_i},
\end{equation}
where $m_i$ is the mass of cell $i$. For a uniform grid, with a constant $\mu$ along the line of sight,
\begin{equation}
    n_M = \frac{\sum_i^{N_{\rm cells}} n_{H_i}^2}{\sum_i^{N_{\rm cells}} n_{H_i}}.
\end{equation}
So, $n_V$ and $n_M$ represent statistical quantities of moments of the density distribution, which would then require to be interpreted through some physical model.

\begin{figure*}[htb!]
    \centering
    \includegraphics[width=0.95\textwidth]{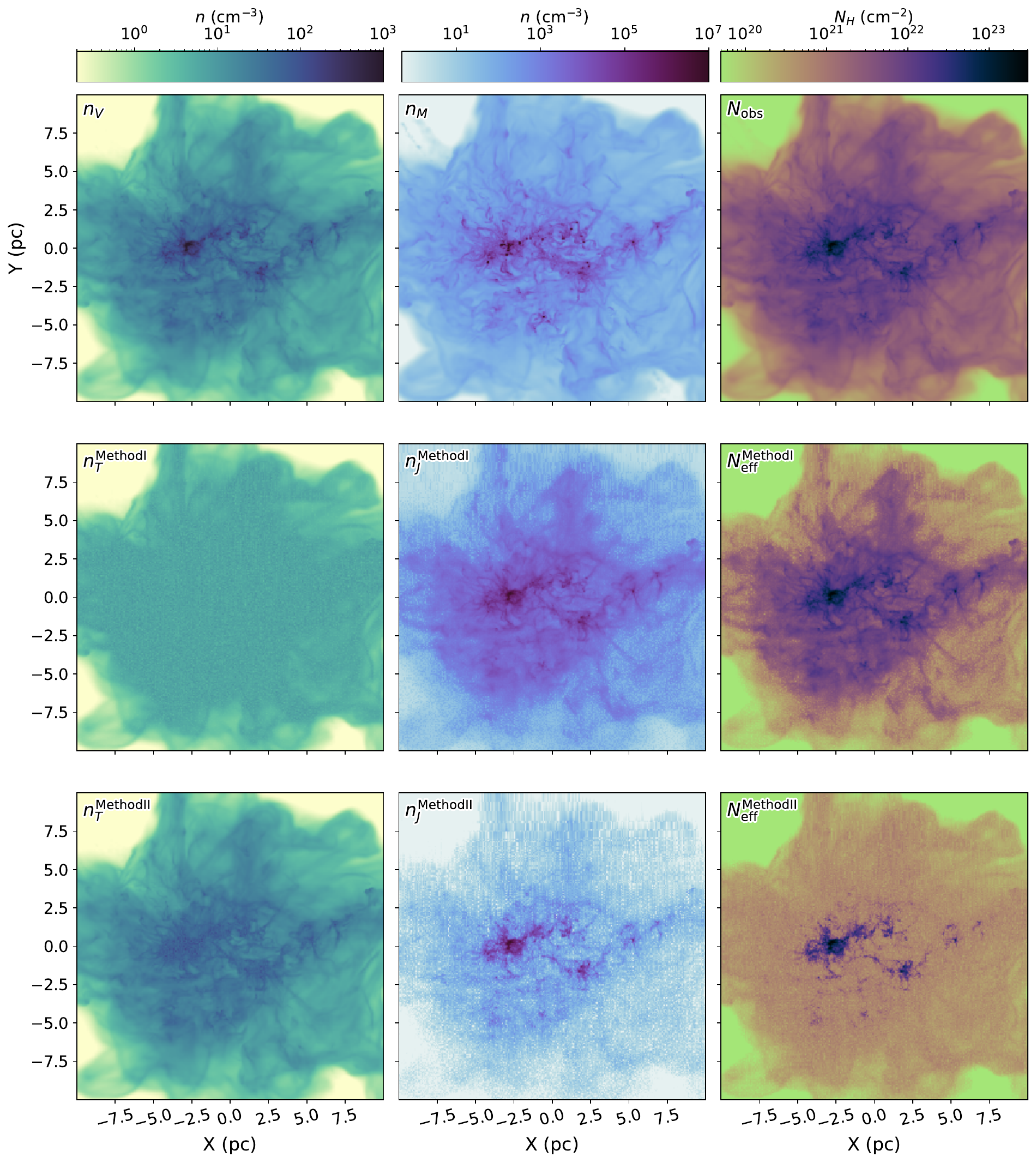}
    \caption{\label{fig:starforge}Summary of {\sc Starforge} simulation comparison results. Top row: line-of-sight volume-weighted average number density (left), line-of-sight mass-weighted average number density (middle), line-of-sight integrated column density (right). Middle row: turbulent number density computed using Method I (left), corresponding gravitational number density (middle), and effective column density (right). Bottom row: same as middle but for Method II.}
\end{figure*}

\begin{figure*}[htb!]
    \centering
    \includegraphics[width=0.95\textwidth]{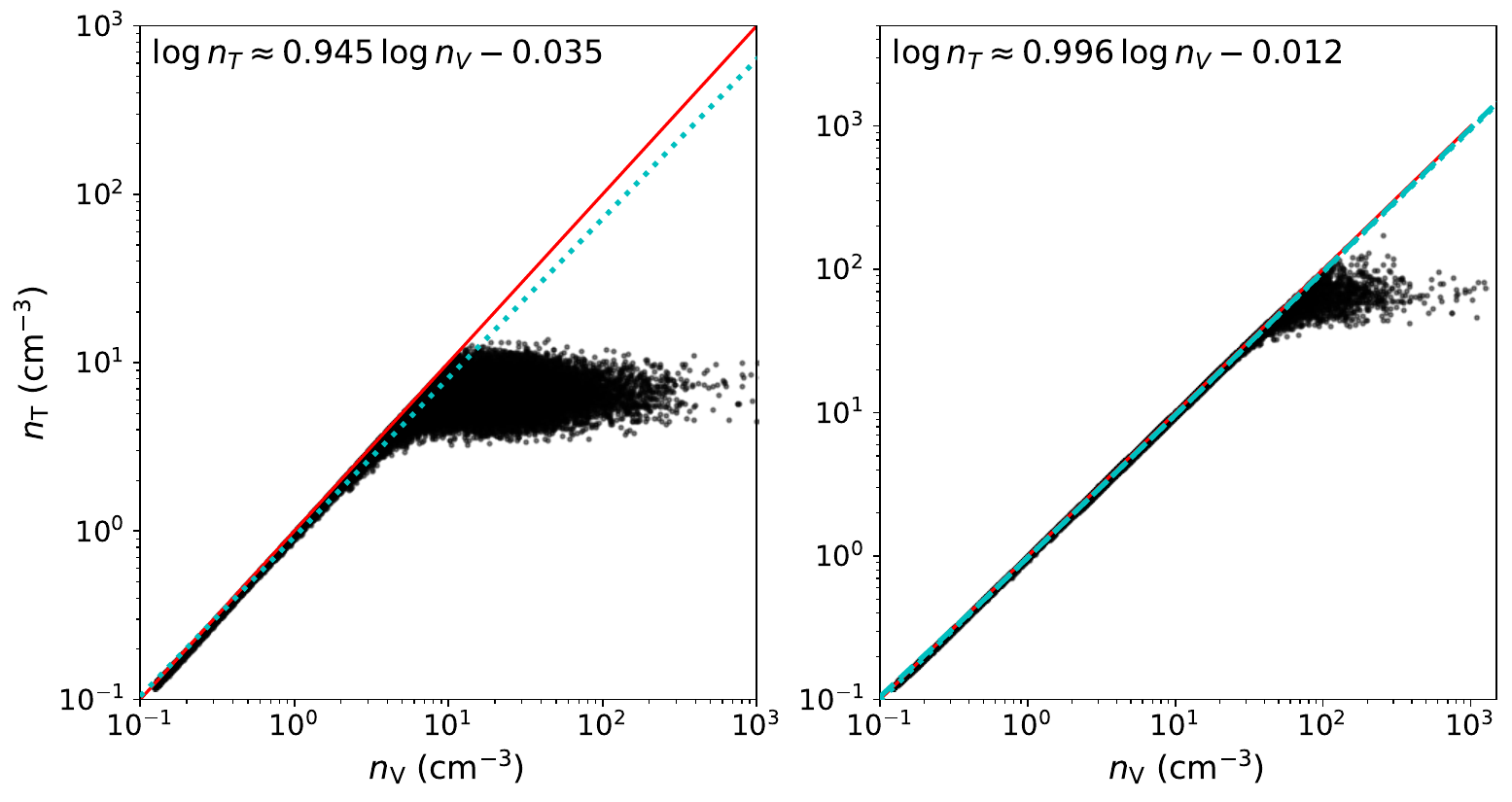}
    \caption{\label{fig:starforge_volume}Derived turbulent densities versus line-of-sight volume-weighted average density using  Method I (left) and Method II (right). The solid red line shows the 1:1 ratios and the cyan dotted line shows the best fit in log-space with the trend annotated in the top left corner.}
\end{figure*}

Figure \ref{fig:starforge} presents the summary of the results comparing the different derived values. In particular, we find that the gravitational density does not directly trace the mass-weighted average density. However, the turbulent density distribution, especially using Method II, seems to qualitatively match the behavior of the volume-weighted average density albeit with an offset. Figure \ref{fig:starforge_volume} shows a comparison of the volume-weighted average density and the computed turbulent density using Methods I and II. The figures present the fits in log-space, with the fits being performed in the density regime before the plateau in $n_T$ for each model, which is equal to $N_0/L$. The Method I turbulent density matches well the volume-weighted average density for low densities, but because of the plateauing featured with this method these two become decoupled. However, there is a tight correlation between the Method II turbulent density and the volume-weighted averaged density for most of the density range, with a slope of unity. This demonstrates that the turbulent density is tracing a volume-averaged line-of-sight density, which is expected since the turbulent envelope around the dense gas is expected to occupy a larger volume than the dense gas itself.

\begin{figure*}
    \centering
    \includegraphics[width=0.95\textwidth]{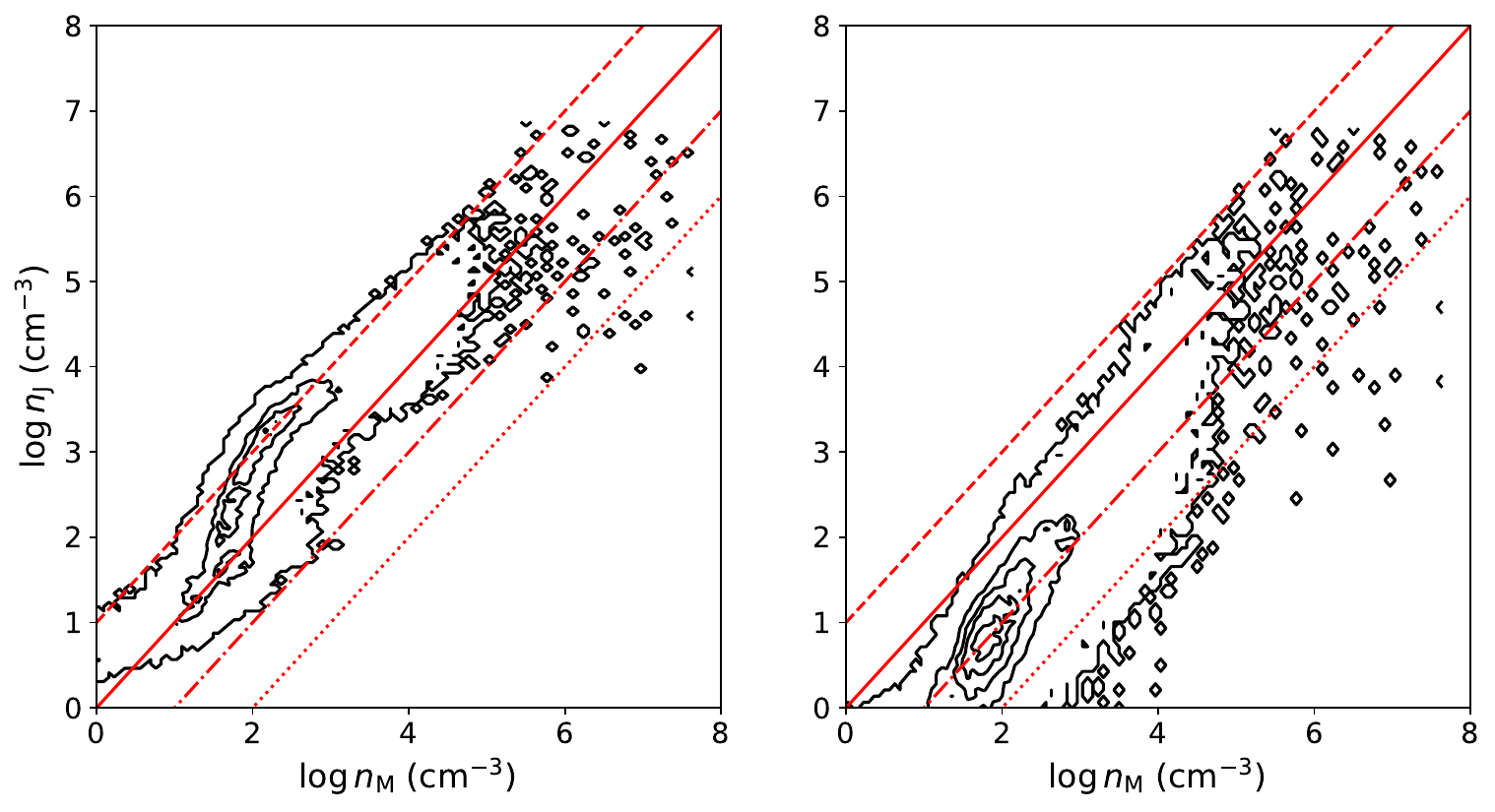}
    \caption{\label{fig:starforge_mass}Derived gravitational densities versus line-of-sight mass-weighted average density using Method I (left) and Method II (right), plotted as contours of the 2D histogram. The red dashed lines show the 10:1 (dashed), 1:1 (solid), 1:10 (dashed-dot), and 1:100 (dotted) ratios.}
\end{figure*}

Figure \ref{fig:starforge_mass} shows the gravitational densities computed using the Method I and II turbulent densities compared to the mass-weighted average density. The Method I gravitational density generally follows between the 10:1 to 1:1 relations against the mass-weighted averages, although there is still a significant spread between the 10:1 and 1:10 trend lines. Method II gravitational density is less than the mass-weighted average by about a factor of 10, although toward higher densities these values start to get closer to 1:1. Despite the significant spread, the two factors appear to be correlated. The additional spread in these correlations could be related to either extra turbulent media being included in the {\sc Starforge} mass-weighted average or the lack of magnetic fields in the physical model -- which the simulations include. We note that we assumed the cloud is isothermal, but the simulations are not isothermal. However, we performed this comparison to best treat the simulations as one could with observational data where spatially variable robust gas temperatures are unavailable. For Method I, the effective column density more closely matches the observed column density while for Method II the effective column density is flatter with a substantial increase toward the highest density regions.

\subsection{Application to Polaris and Taurus}
\begin{figure*}
    \centering
    \includegraphics[width=0.48\textwidth]{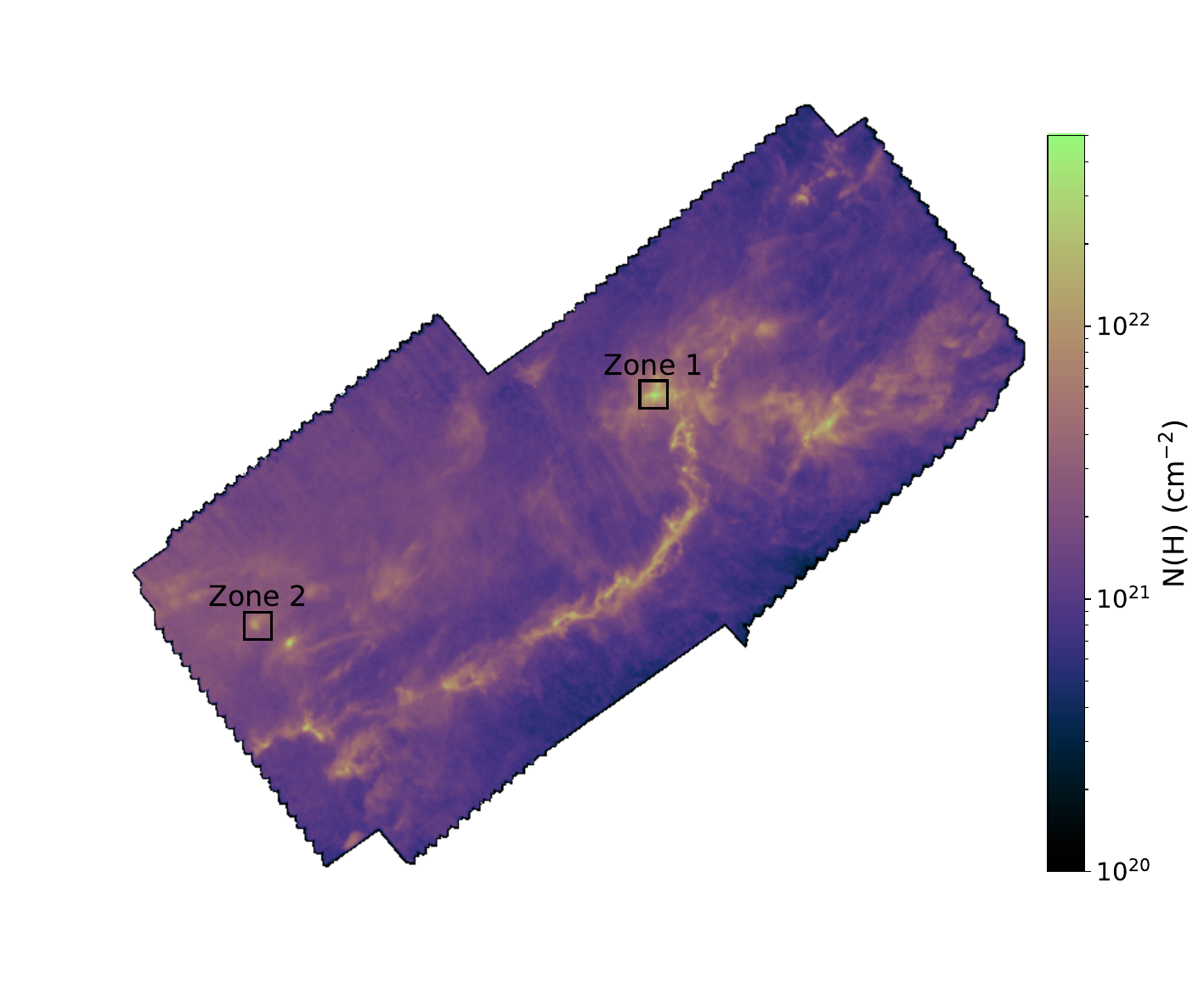}
    \includegraphics[width=0.48\textwidth]{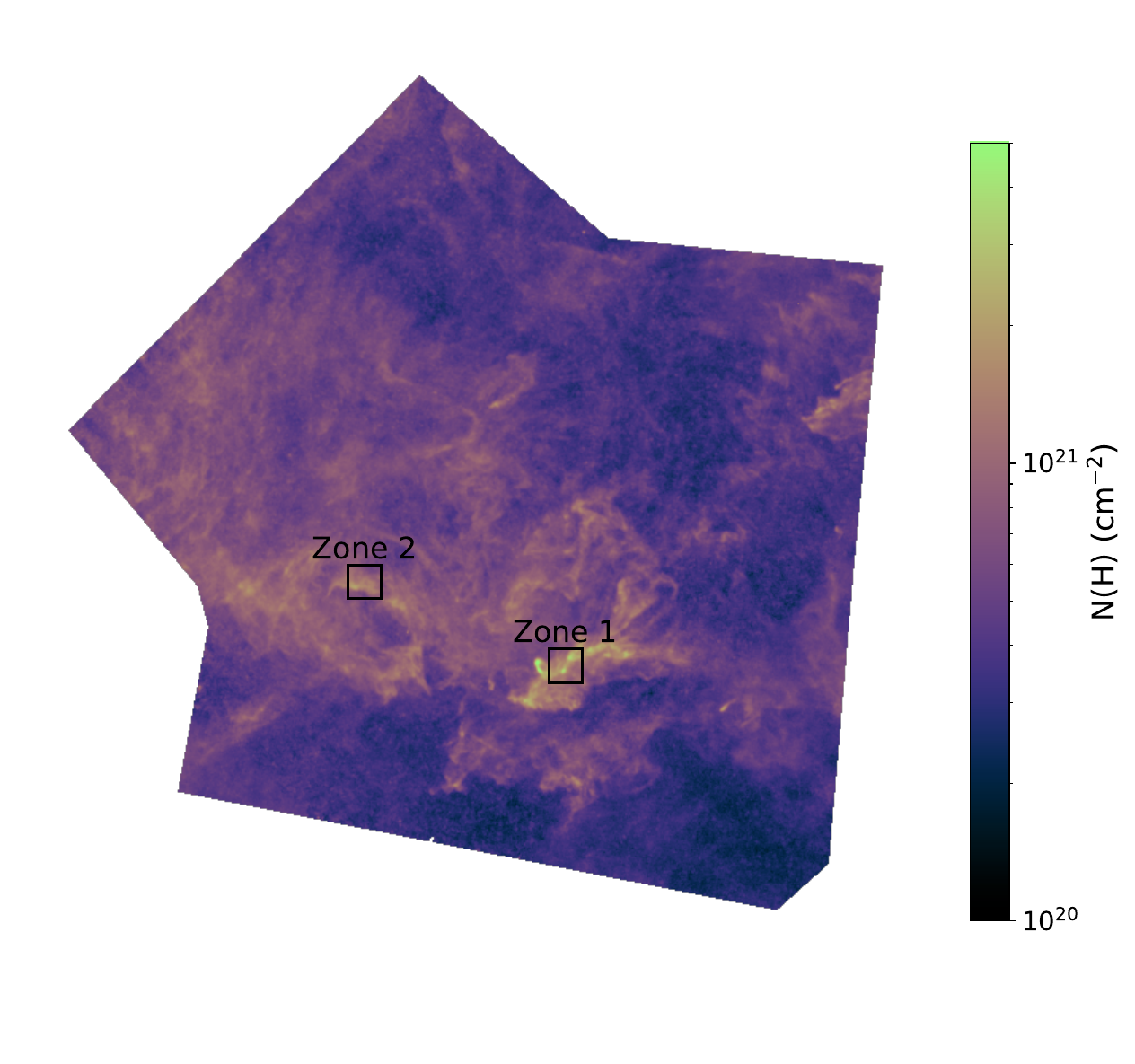}
    \caption{\label{fig:column}Column density maps for Taurus (left) and Polaris Flare (right) from the Herschel Gould Belt Survey. Annotated are two subregions, Zone 1 and Zone 2 which are examined in this work.}
\end{figure*}

Within the Milky Way, there is a vast range of different morphologies of GMCs, from diffuse condensations (Polaris Flare), low-mass star-forming regions (Taurus Molecular Cloud), high-mass star-forming regions (Orion Molecular Cloud), and quiescent, dense molecular clouds, such as infrared dark clouds (IRDCs) and clouds in the galactic center, such as The Brick molecular cloud \citep[see reviews by][]{Tan2014, Zucker2023, Henshaw2023}. Understanding the dynamical and chemical evolution of GMCs is paramount to our understanding of star formation. We apply our model to two well-studied molecular clouds, Taurus and the Polaris Flare. The column density maps for the Taurus L1495/B213 filament \citep{Kirk2013, Palmeirim2013, Marsh2016} (region N3 in \citet{Kirk2013}) and Polaris Flare \citep{Men'shchikov2010} are taken from the Herschel Gould Belt Survey \citep{Andre2010}. Figure \ref{fig:column} shows the column density maps for Taurus and the Polaris Flare, along with two subregions we investigate using the full bivariate probability distributions (Method III).

\subsubsection{Taurus}
\begin{figure}
    \centering
    \includegraphics[width=\columnwidth]{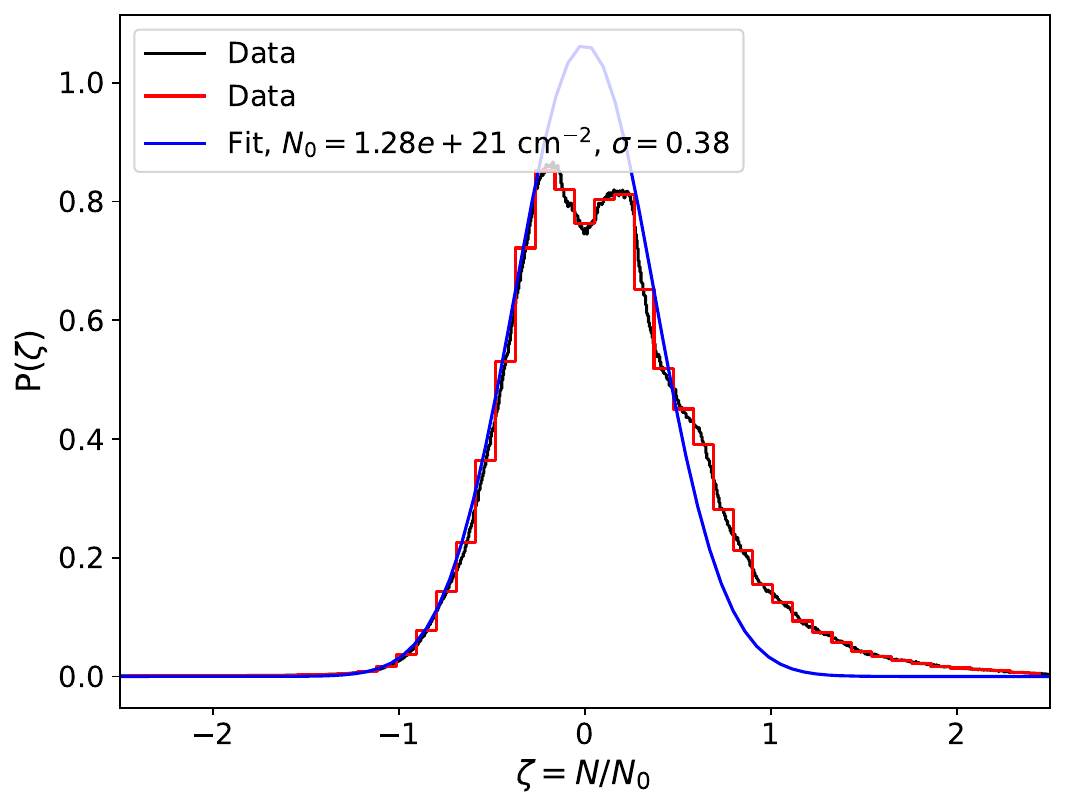}
    \caption{\label{fig:hist_taurus} N-PDF for the Taurus cloud with the best fit log-normal distribution. The black and red lines show the histograms of the data, where the black uses an automated procedure for binning and the red uses a fixed bins more typical to previous studies. The blue line shows the log-normal utilized in this work.}
\end{figure}

The Taurus molecular cloud provides a good test for this method, since it has a combination of both dense and diffuse regions, with the dense gas being enhanced toward the well-characterized L1495/B213 region \citep{Lombardi2010}. Taurus has ongoing low-mass star formation and, since it is nearby with a distance of 140 pc \citep{Schlafly2014}, has been extensively studied. The N-PDF for Taurus is shown in Figure \ref{fig:hist_taurus} along with the lognormal that we used to model the turbulence. While the distribution shows a double-peaked structure, the log-normal that we use to fit for the turbulent component of the N-PDF reproduces the lower column density gas well. The double-peaked structure seems to be resulting from the turbulent diffuse envelope being marginally different between the filament and the northwestern region with the smaller cores, thus picking out two more separate large-scale features. The double-peaked feature largely goes away when the N-PDF histogram is binned with the number of bins used in previous studies \citep[][]{Marsh2016, Schneider2022}, however the remnant of the double peak likely means that the regions have slightly different turbulence properties. We use a molecular cloud thickness of 1 pc for Taurus \citep{Qian2015}. This width is consistent with the Plummer model fits from \citet{Zucker2021} of a flattening radius of $1.2^{+0.3}_{-0.3}$ pc, while their two-component Gaussian fits gave widths of $2.3^{+0.1}_{-0.1}$ pc and $11.9^{+0.6}_{-0.5}$ pc tracing a more dense and diffuse component, respectively. We assume a gas temperature of 15 Kelvins, which is greater than the kinetic temperature measured in the dense gas by ammonia \citep{Seo2015}, but consistent with dust temperatures measured in the region ranging between 10 - 15 Kelvins \citep{Kirk2013}. However, these small temperature differences do not noticeably alter the results. The fit results in an average Mach number of $M_s = 2.4$, resulting in a turbulence velocity dispersion of approximately $0.7$ km s$^{-1}$ and a transition density $n_{\rm tr} \approx 6\times10^4$ cm$^{-3}$.

\begin{figure*}
    \centering
    \includegraphics[width=0.48\textwidth]{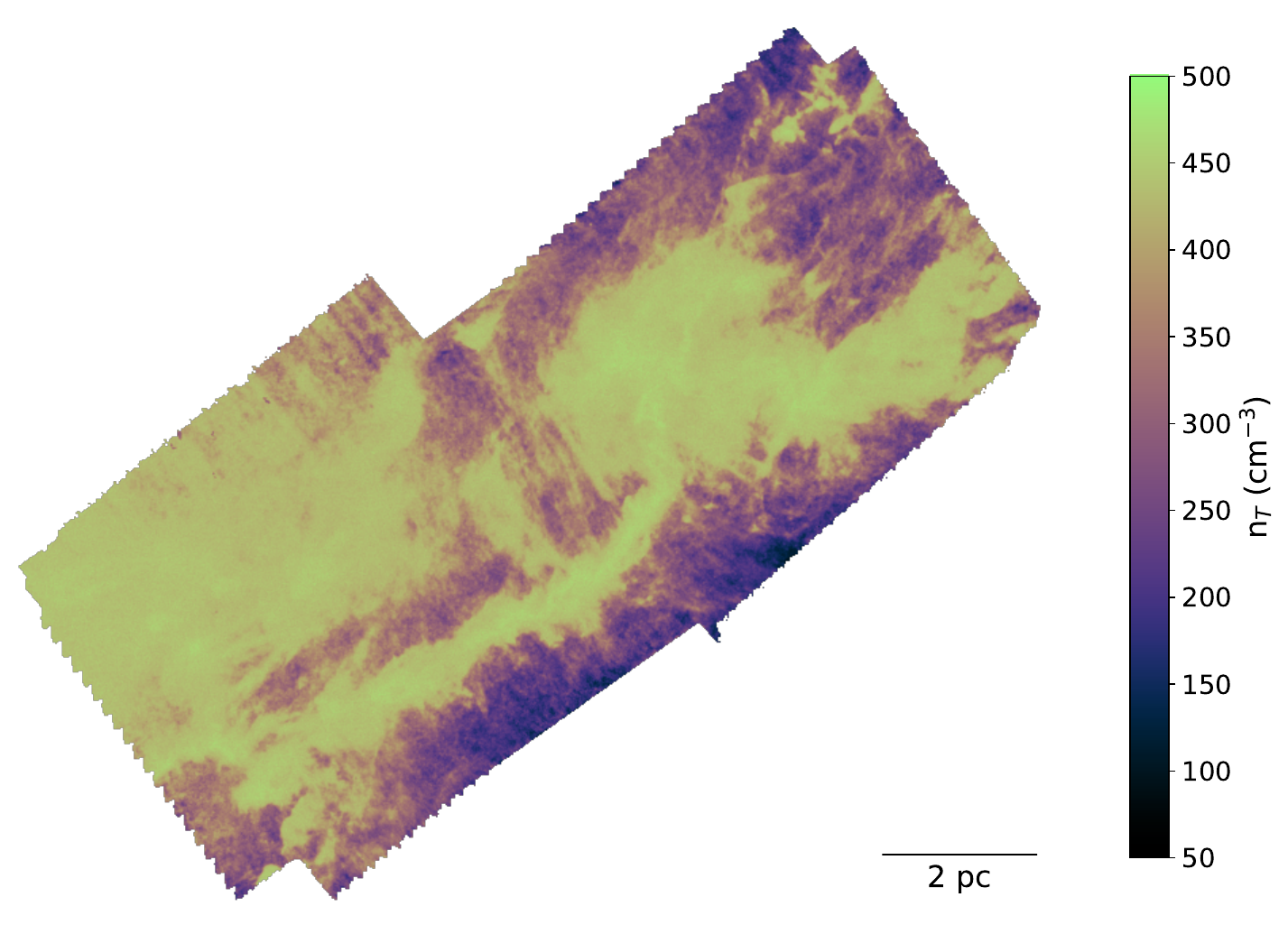}
    \includegraphics[width=0.48\textwidth]{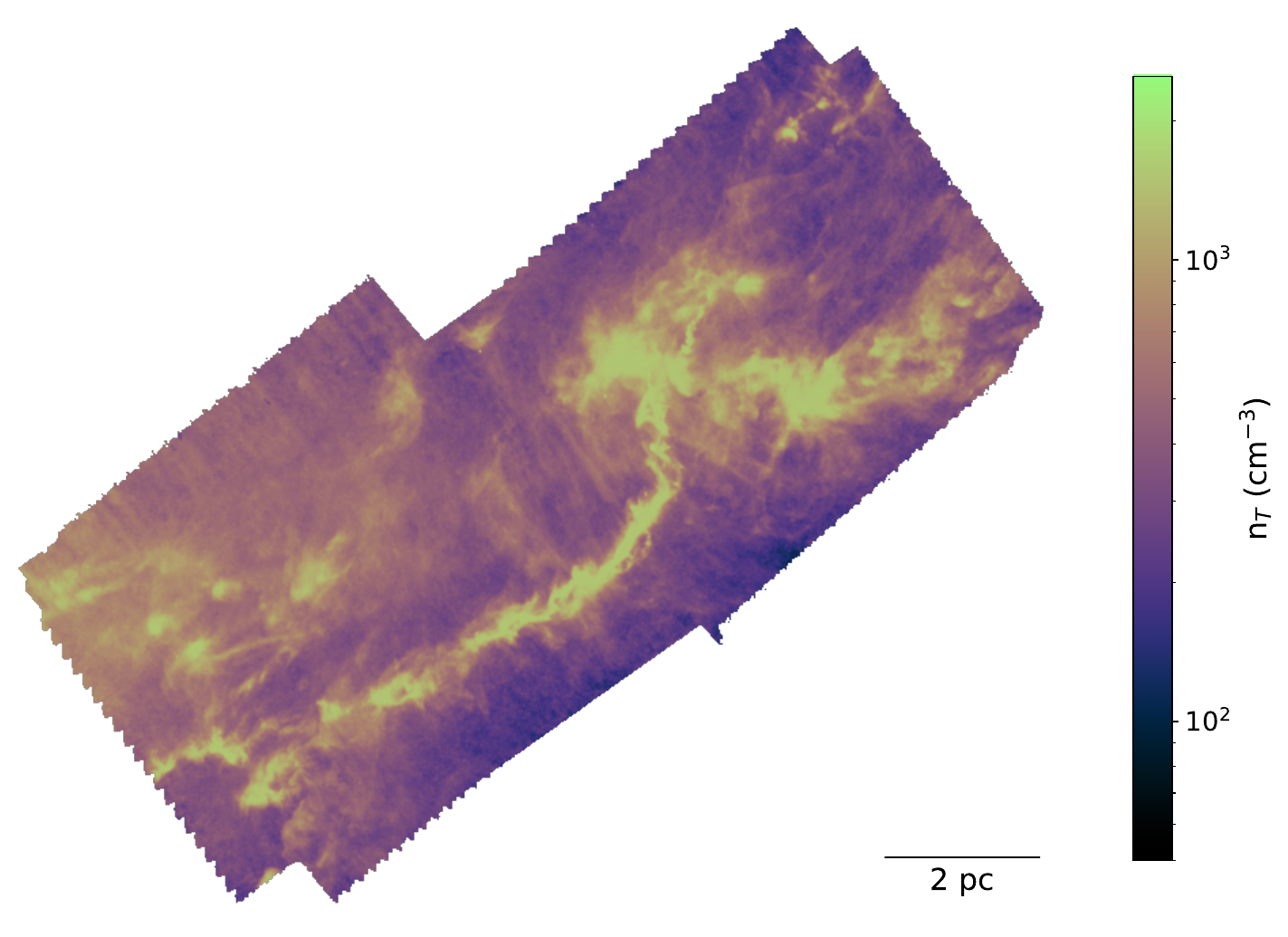}
    \caption{\label{fig:taurus_turb} Derived turbulent gas density for Taurus using Method I(left) and Method II (right).}
\end{figure*}

\begin{figure*}
    \centering
    \includegraphics[width=0.48\textwidth]{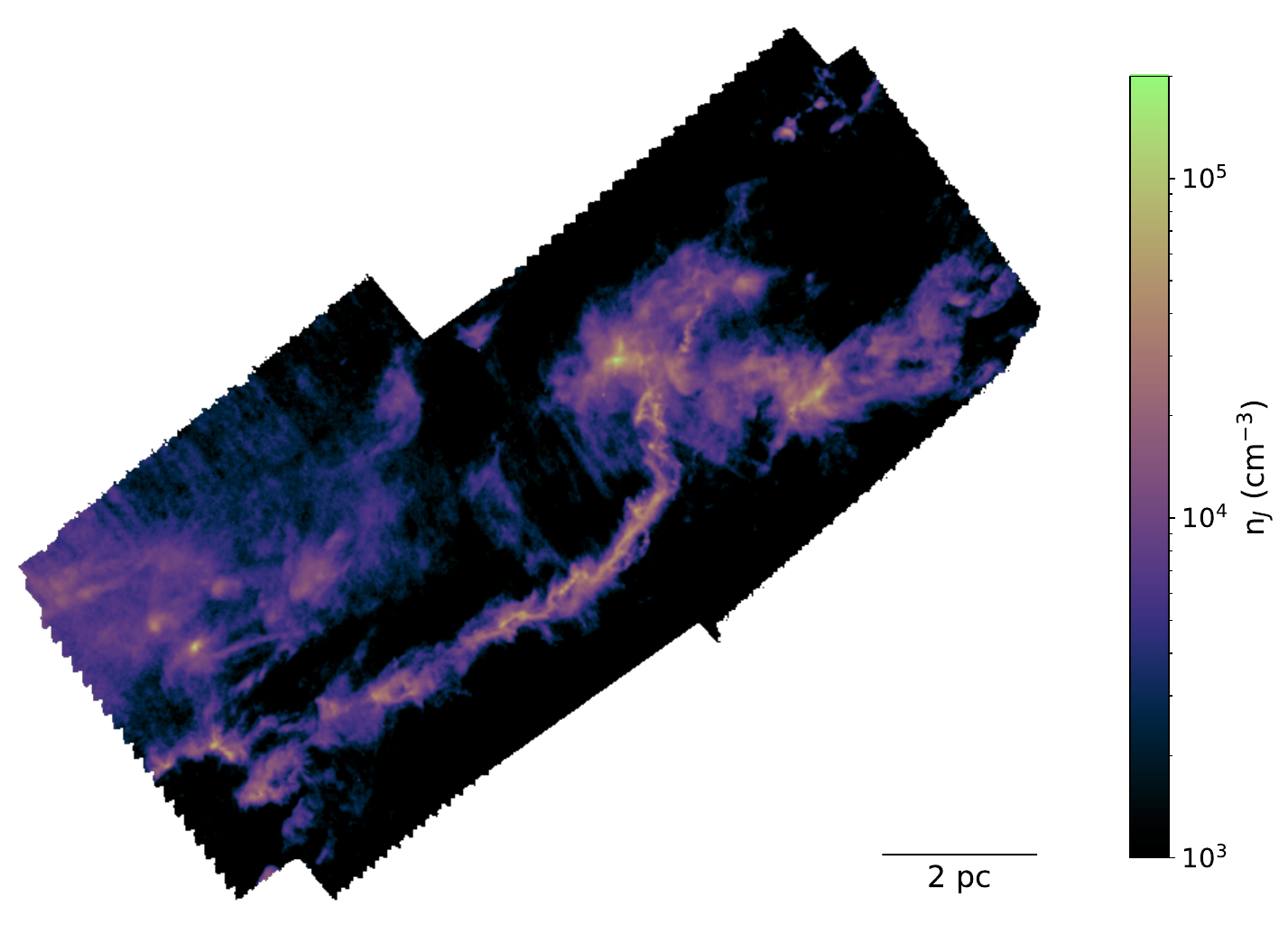}
    \includegraphics[width=0.48\textwidth]{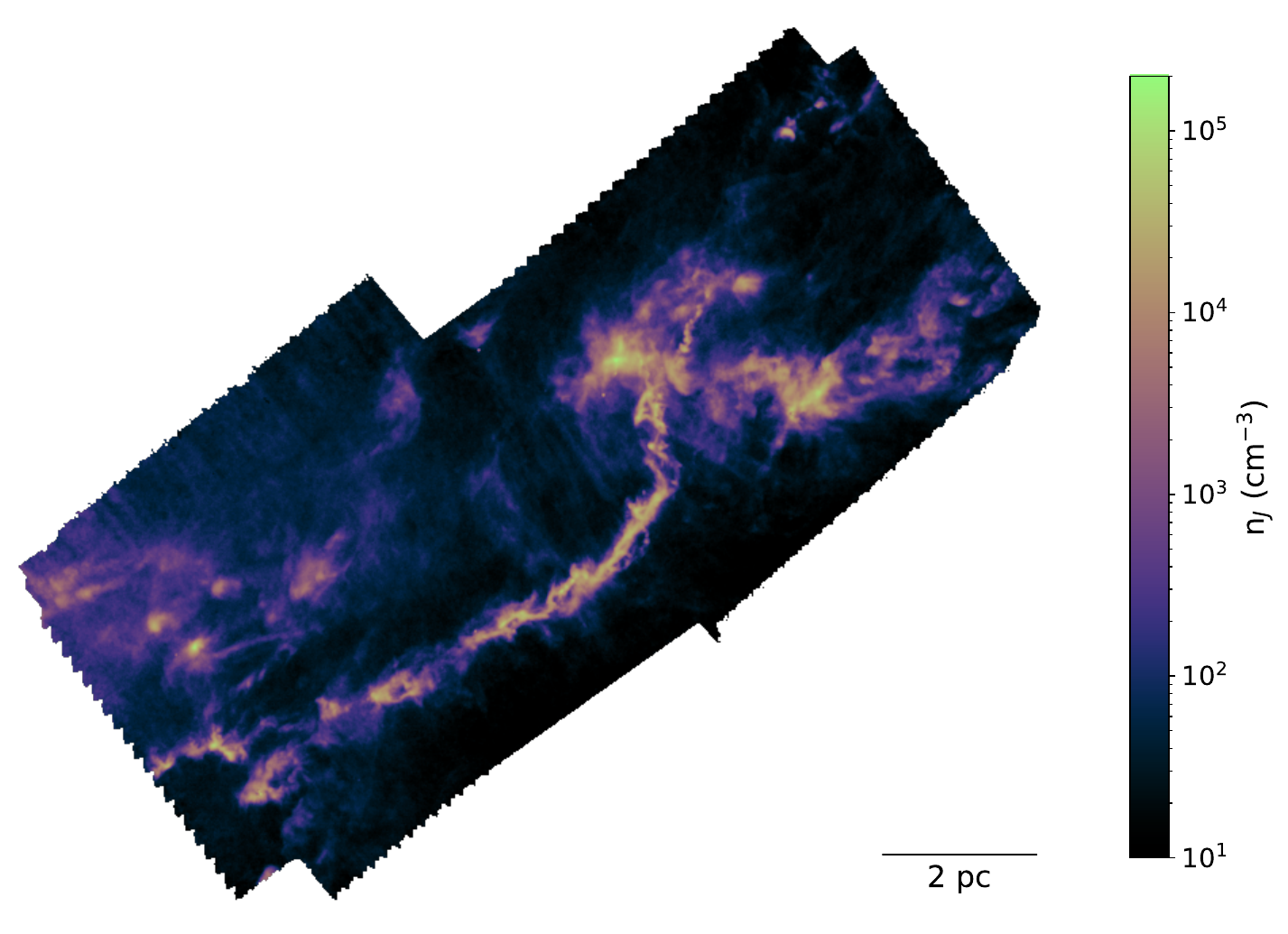}
    \caption{\label{fig:taurus_grav} Derived gravitational density for Taurus using Method I (left) and Method II (right).}
\end{figure*}

\begin{figure*}
    \centering
    \includegraphics[width=0.48\textwidth]{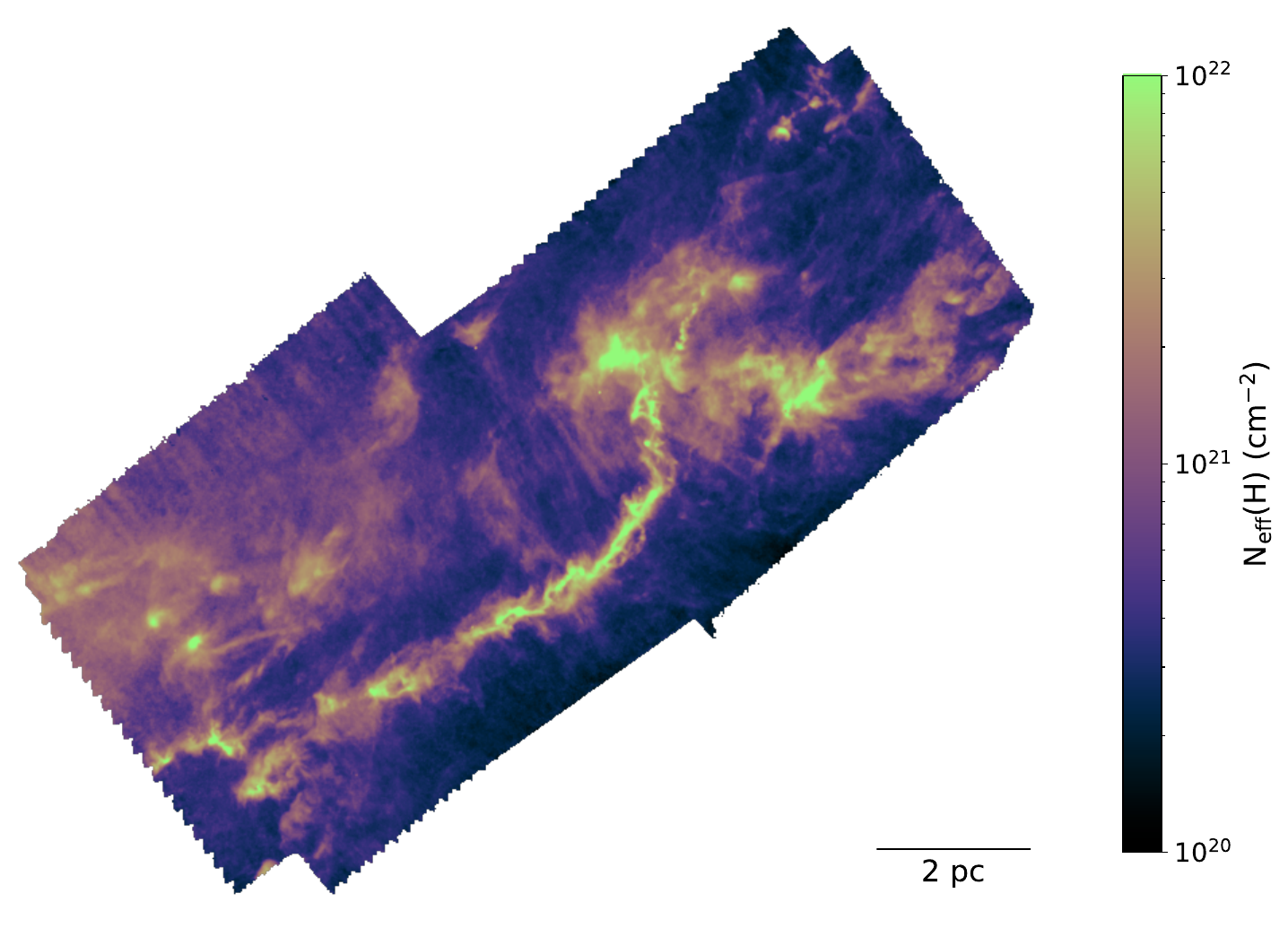}
    \includegraphics[width=0.48\textwidth]{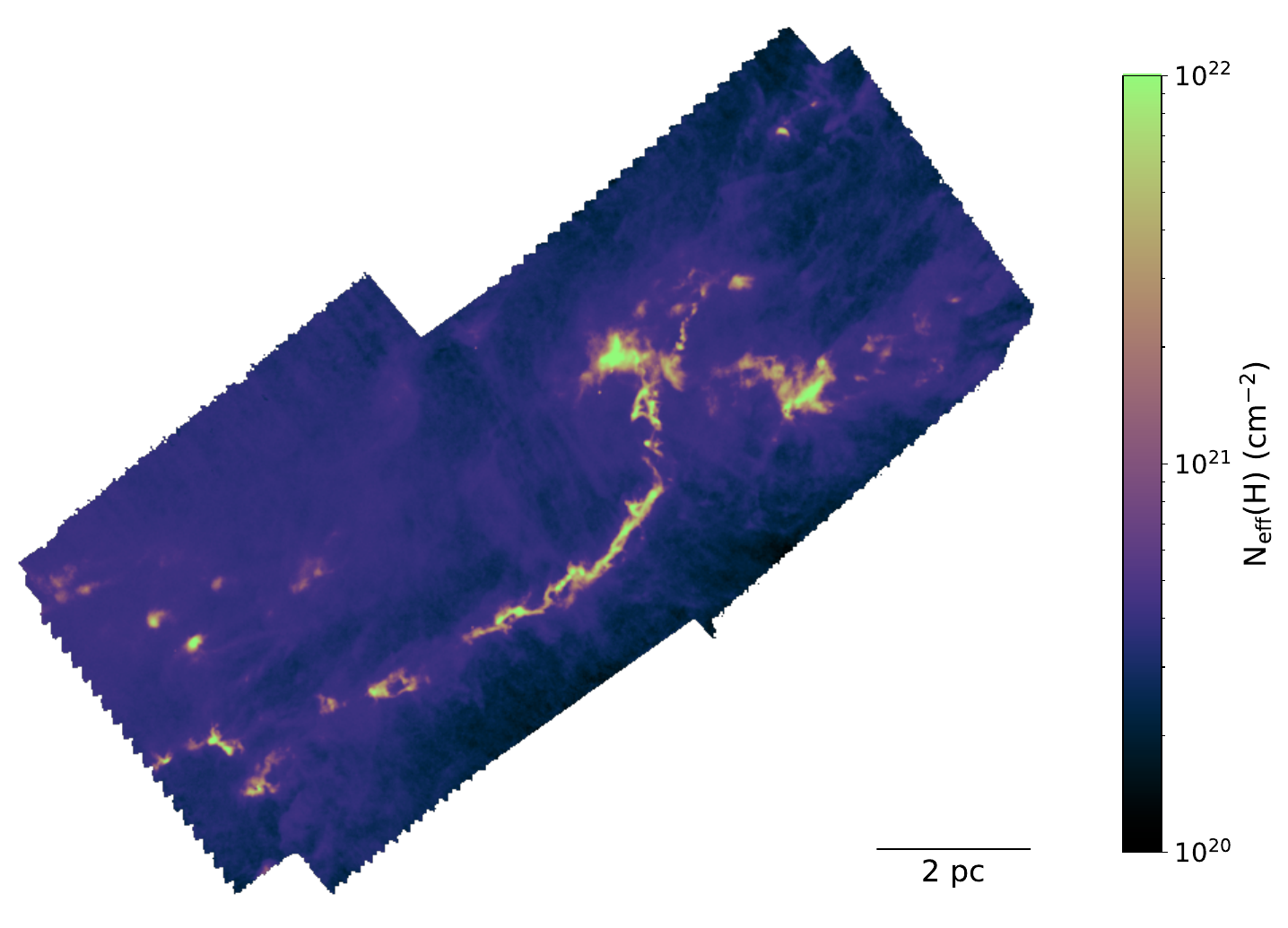}
    \caption{\label{fig:taurus_neff} Derived effective column densities for Taurus using Method I (left) and Method II (right) of the turbulent column density distribution along each line of sight.}
\end{figure*}

Figure \ref{fig:taurus_turb} shows the derived turbulent gas density across the Taurus molecular cloud using both Method I and II from the sampled values of $N_T$. As expected, the Method I values plateau at $n_T \approx 450$ cm$^{-3}$ which corresponds roughly to the peak value of the N-PDF divided by 1 pc. The Method II value of $n_T$ roughly traces the column density structure of the cloud peaking at $n_T \approx 2\times 10^3$ cm$^{-3}$ in the densest regions. Figure \ref{fig:taurus_grav} shows the gravitational density, $n_J$, throughout the cloud. As expected, the dense gas is primarily situated in the filament peaking to $n_J \approx 10^5$ cm$^{-3}$. Both approximations for the turbulent column density return similar results for the gravitational density in the densest regions, although there is a significantly larger range when using Method II, decreasing the gravitational density in the diffuse regions to $n_J < 100$ cm$^{-3}$. These results, when convolved to a lower resolution, are consistent with the results of \citet{Li2012} and \citet{Xu2023}, who both predict an average filament density of $1-4\times 10^{4}$ cm$^{-3}$, tracing factors of 5-10 lower spatial resolutions. A further benefit of the method is the ability to predict the effective column density. Figure \ref{fig:taurus_neff} shows the effective column density for the two methods for the dense gas component. Now, there is a stark difference between the two different assumptions with the use of the Method II $N_T$ resulting in a cloud in which only the filament has a noticeably high effective column density. By construction, we require that the effective column density is always less than the observed column density. This is demonstrated in Figure \ref{fig:neff_nobs_taurus}, which shows the ratio of the effective column density computed with Model I against the observed column density. Figure \ref{fig:neff_nobs_taurus} shows that in the diffuse regions, it limits to a fraction of the observed column density, while in the dense regions it approaches the observed column density since the attenuating column density is dominated by the gravitational factor. The significant difference in the diffuse gas between the effective and the observed column densities comes about because the effective column density is the attenuating column density from the gravitational component out to the cloud boundary and this component (the blue box in Figure \ref{fig:schematic}), may be randomly anywhere along the line of sight. Further, since the turbulence component dominates here, for Method I, the maximal value then be half the turbulence column density, further reduced by the random depth factor, $\mathcal{R}$ in Equation \ref{eq:NTeff}.

\begin{figure*}
    \centering
    \includegraphics[width=0.48\textwidth]{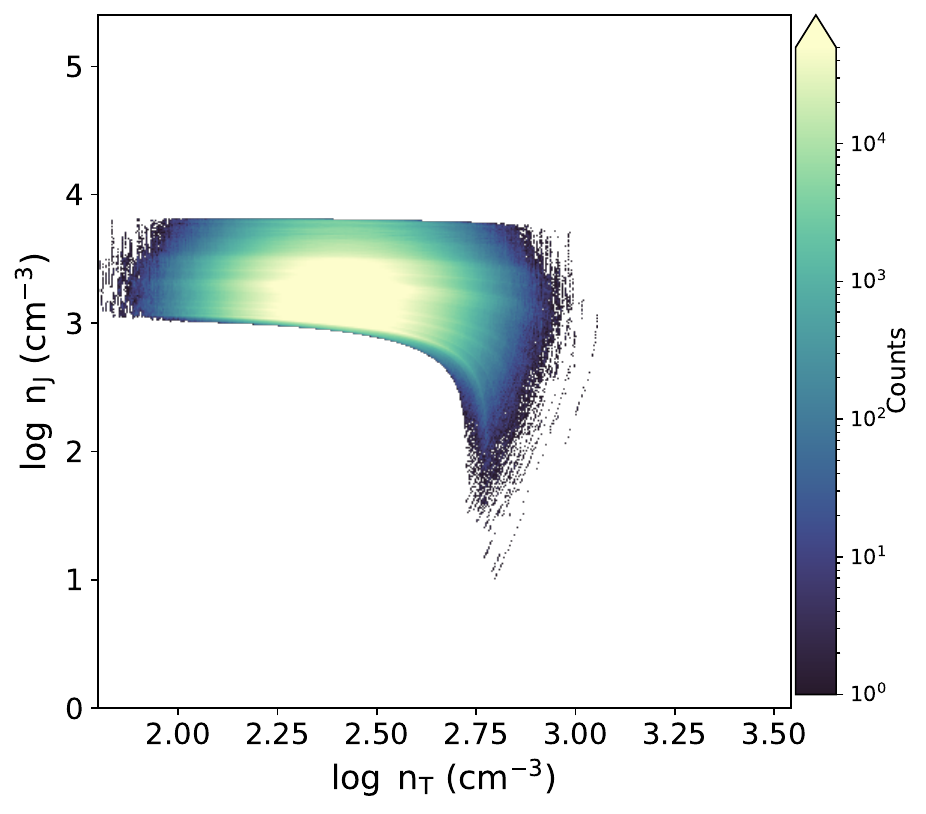}
    \includegraphics[width=0.48\textwidth]{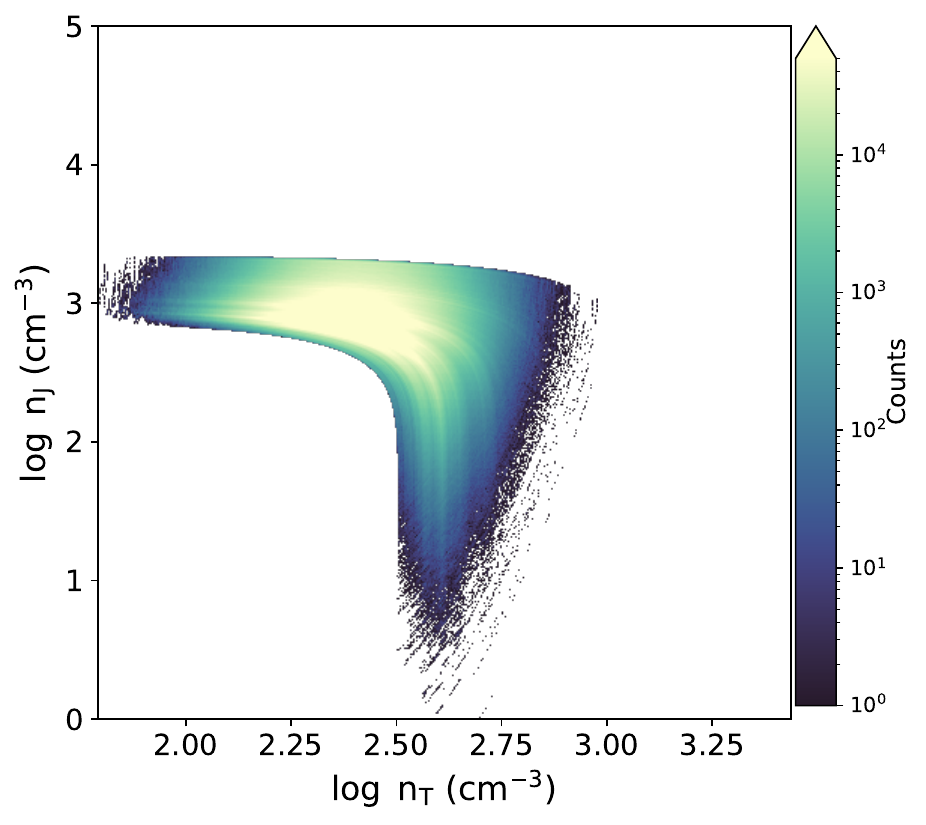}
    \caption{\label{fig:taurus_bivariate} Method III bivariate distributions of the turbulent and gravitational densities for Taurus in Zone 1 (left) and Zone 2 (right).}
\end{figure*}

\begin{figure}
    \centering
    \includegraphics[width=\columnwidth]{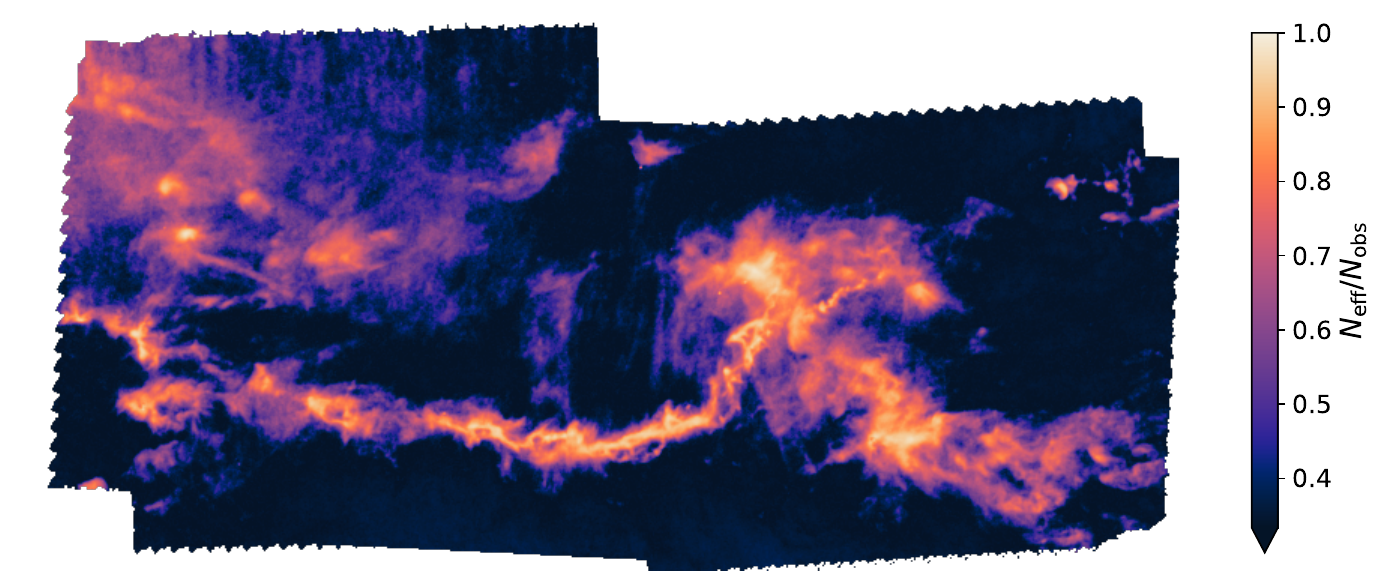}
    \caption{\label{fig:neff_nobs_taurus} Ratio of the effective column density to the observed column density for Taurus using Method I.}
\end{figure}

Finally, Figure \ref{fig:taurus_bivariate} shows the bivariate distributions of $n_T$ and $n_J$ for the two subregions indicated in Figure \ref{fig:column} from Method III. Due to the definition that $N_{\rm obs} = N_T + N_J$ the distribution is highly correlated. Zone 1 is centered around a high-density knot within the filament complex while Zone 2 is situated within the more diffuse region. This is reflected in the bivariate distributions: while both have similar ranges for $n_T$, Zone 1 has a higher overall $n_J$ than Zone 2 by nearly an order of magnitude. Further, the lower column density environment traced by Zone 2 leads to a significant tail in the distribution toward low gravitaitonal masses, such that $n_J < n_T$ along lines of sight. The bivariates in Figure \ref{fig:taurus_bivariate} also demonstrate where Methods I and II are most sensitive: Method I traces the peak of the bivariate distribution, while Method II is tracing the high $n_T$ boundary edge of the bivariate distribution. Compared to the results using Method I and Method II, the bivariate distribution provides the complete information of the probable turbulent and gravitational densities along the line of sight rather than characteristic values for each component. The bivariate could be utilized to generate line emmissities or probability distributions of important physical quantities, such as the free-fall time, for each pixel in a column density map, rather than a singular value. The benefit of this is a more comprehensive exploration of a range of physics or chemistry, and will be investigated in future work. However, there is a significant memory overhead for storing the underlying samples.

\subsubsection{Polaris Flare}
\begin{figure}
    \centering
    \includegraphics[width=\columnwidth]{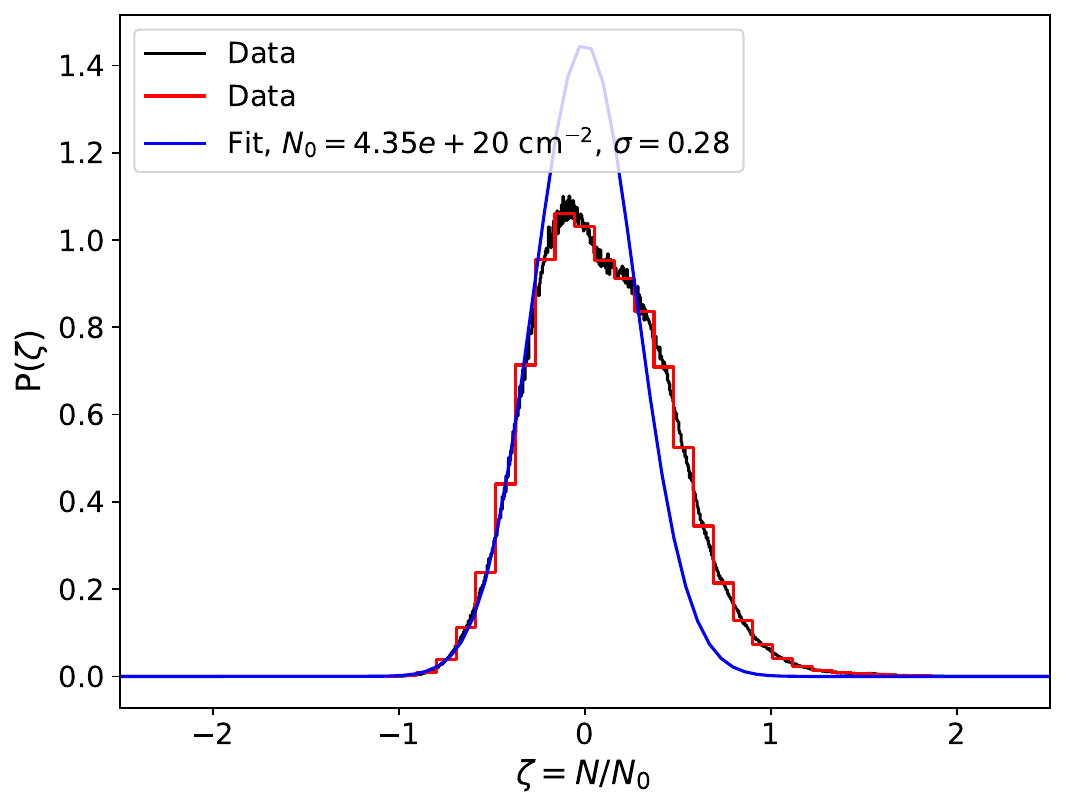}
    \caption{\label{fig:hist_polaris} Same as Figure \ref{fig:hist_taurus} but for Polaris.}
\end{figure}

The Polaris Flare is a high-latitude diffuse cloud. The Polaris Flare exhibits little to no signs of ongoing star formation, with 5 possible bound prestellar cores found by \citet{Andre2010, Ward-Thompson2010}, and is characterized by incredibly diffuse gaseous structure and striations. The distance to Polaris was once highly debated, with distances between 100 and 400 pc argued (see \citet{Schlafly2014}), but new results using Gaia data put the distance more firmly around 340 - 360 pc \citep{Zucker2020}. Further, it has been argued that magnetic field threading the cloud is dynamically important \citep{Panopoulou2016, Skalidis2023}. Figure \ref{fig:hist_polaris} shows the N-PDF for Polaris and the utilized log-normal turbulent distribution. The distribution is peaked at a much lower column density than Taurus with a thinner log-normal Gaussian. Since there is no measured cloud thickness, we assume a thickness of 2.5 pc, similar to other quiescent or low-mass star-forming regions \citep{Qian2015, Zucker2021}. We assume a gas temperature of 15 Kelvins, and the resulting fit Mach number of $M_s = 1.55$ results in a turbulent velocity dispersion of approximately 0.46 km s$^{-1}$ and a transition density, $n_{\rm tr} \approx 1750$ cm$^{-3}$.

The differences between Method I and II $n_T$ are similar to that of Taurus. Figure \ref{fig:polaris_turb} shows that for turbulent density, the Method I $n_T$ plateaus to approximately $n_T \approx 60$ cm$^{-3}$ while for Method II the turbulent density can exceed 200 cm$^{-3}$. The gravitational densities, shown in Figure \ref{fig:polaris_grav}, are two orders of magnitude lower than that of Taurus, and the Method II value of $n_J$ primarily highlights the feature dubbed the ``saxophone'' \citep{Schneider2013}. Further, Figure \ref{fig:polaris_neff} shows that the effective column densities are primarily $N_{\rm eff} < 10^{21}$ cm$^{-2}$, resulting in extinctions less than unity. These results agree with the fact that the Polaris Flare is a diffuse cloud, although still molecular.

Figure \ref{fig:polaris_bivariate} shows the Method III bivariate distributions of $n_T$ and $n_J$ for the two subregions indicated in Figure \ref{fig:column}. Zone 1 is centered within the saxophone and Zone 2 is within a more diffuse region. These regions exhibit similar bivariate distributions, with only a factor of a few difference in the peak values of $n_J$. For Zone 1, the distribution peaks around $\log n_T = 1.65$ and $\log n_J = 2.0$, while for Zone 2, the distribution peaks around $\log n_T = 1.6$ and $\log n_J = 1.8$. The similarity of these bivariates is primarily because Polaris Flare, even in the more dense regions, is still primarily a diffuse cloud dominated by the turbulent envelope, with the peak $n_J$ in the map only slightly exceeding 1000 cm$^{-3}$, two orders of magnitude lower than Taurus.

\subsection{Use case: carbon cycle emission predictions for Taurus}
As an example use case for the model results, we use the above densities and the effective column density to produce synthetic emission maps for important Carbon cycle lines [C II] (1900 GHz), [C I] (492 GHz) and CO (J = 1-0) (115 GHz). The line brightness temperatures are derived from photo-dissociation region models using the code {\sc 3d-pdr}\footnote{\url{https://uclchem.github.io/3dpdr/}} \citep{Bisbas2012}. We used a reduced network, a subset of the UMIST 2012 gas-phase chemical network \citep{McElroy2013} which has 33 species and 330 reactions with standard ISM abundances at solar metallicity \citep{Rollig2007}. {\sc 3d-pdr} solves for the thermal balance and level populations of key coolants and evolves the chemical evolution to steady state. The models consist of a grid of extinctions, from $0 \leq A_V \leq 25$ and hydrogen nuclei number densities, $1 \leq n_H \leq 10^6$, irradiated with a far-ultraviolet field of the strength of 1 Draine field \citep{Draine1978} and a total H$_2$ cosmic-ray ionization rate of 10$^{-16}$ s$^{-1}$ \citep{Neufeld2017}. The brightness temperatures were previously used in \citet{Bisbas2019} (see for details). These models are not meant to represent a sophisticated three-dimensional treatment of the cloud but instead can be utilized to rapidly estimate the brightness temperature of specific lines throughout the region. Synthetic emission maps derived through interpolating across grids of astrochemical models using the estimated density and effective column density will be very useful for interpreting observations and for predicting line fluxes for future observations.

\begin{figure*}
    \centering
    \includegraphics[width=0.32\textwidth]{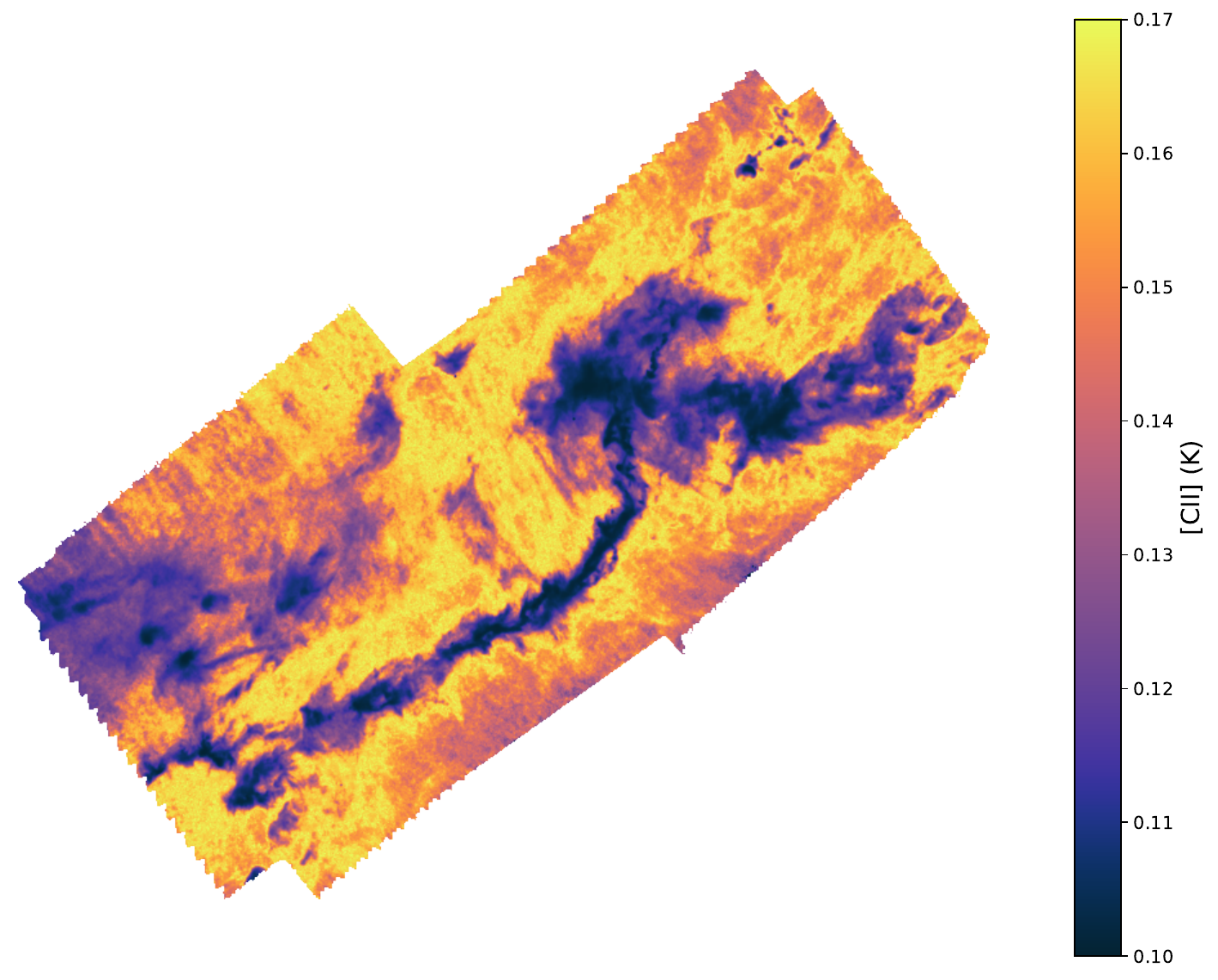}
    \includegraphics[width=0.32\textwidth]{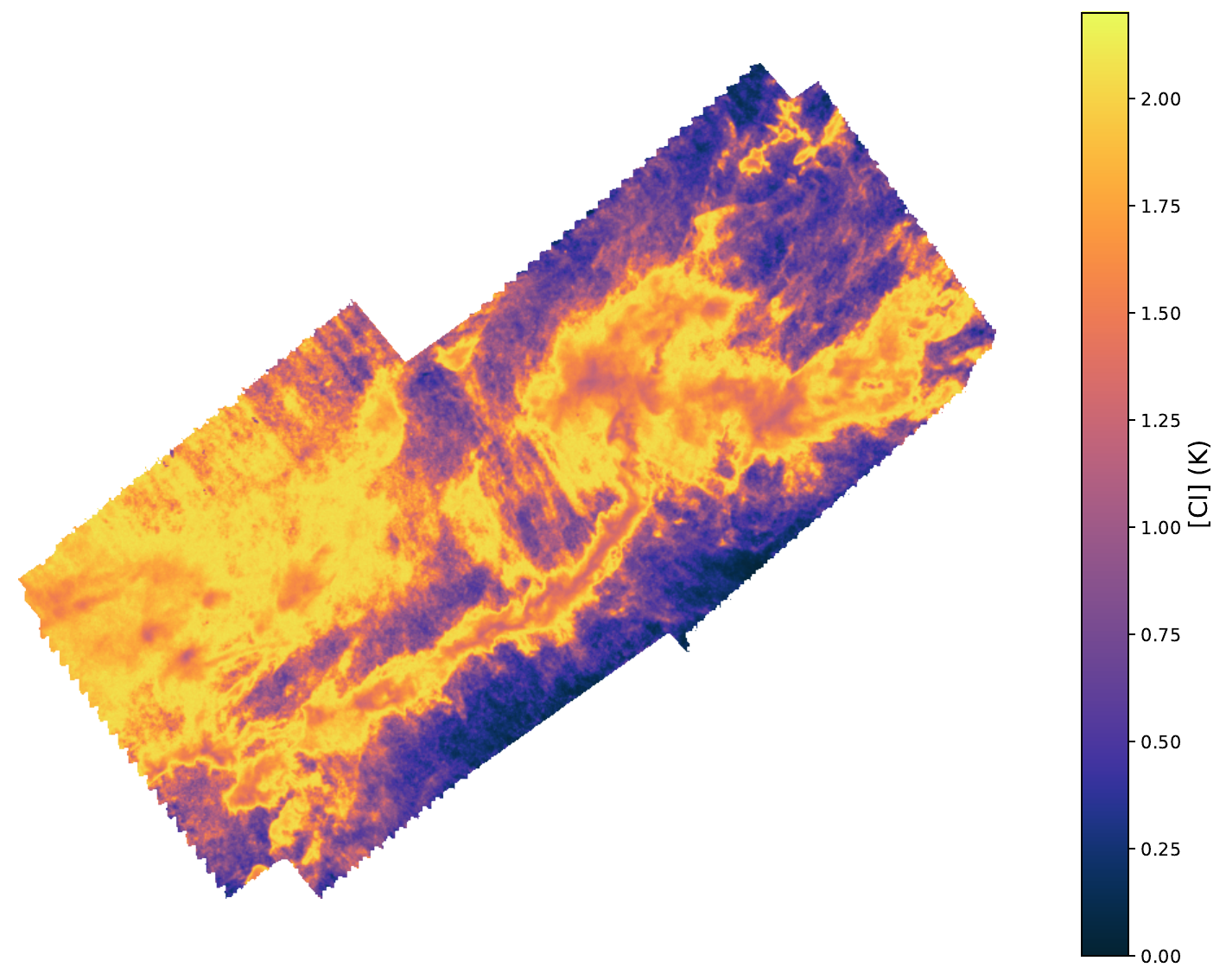}
    \includegraphics[width=0.32\textwidth]{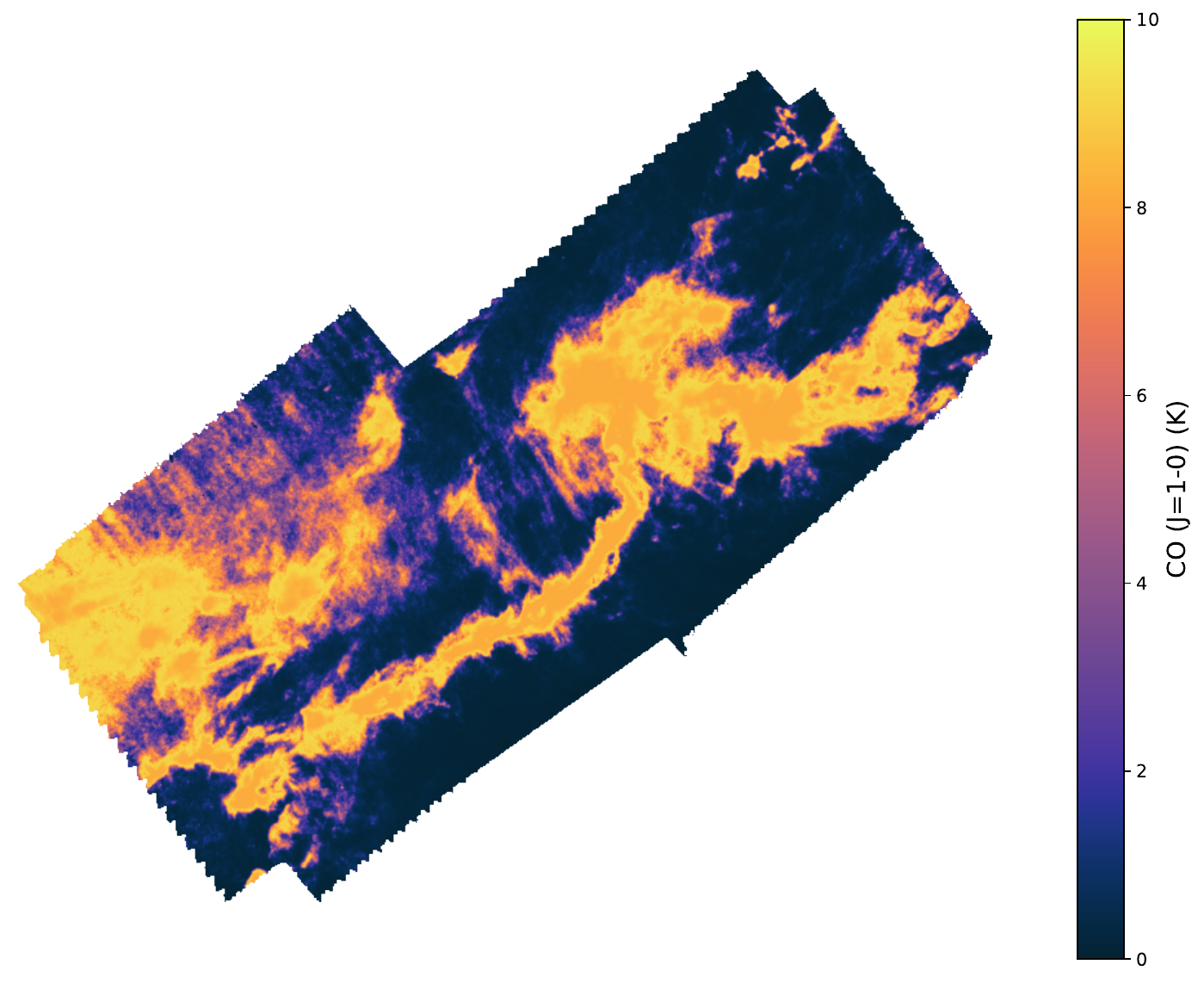}
    \includegraphics[width=0.32\textwidth]{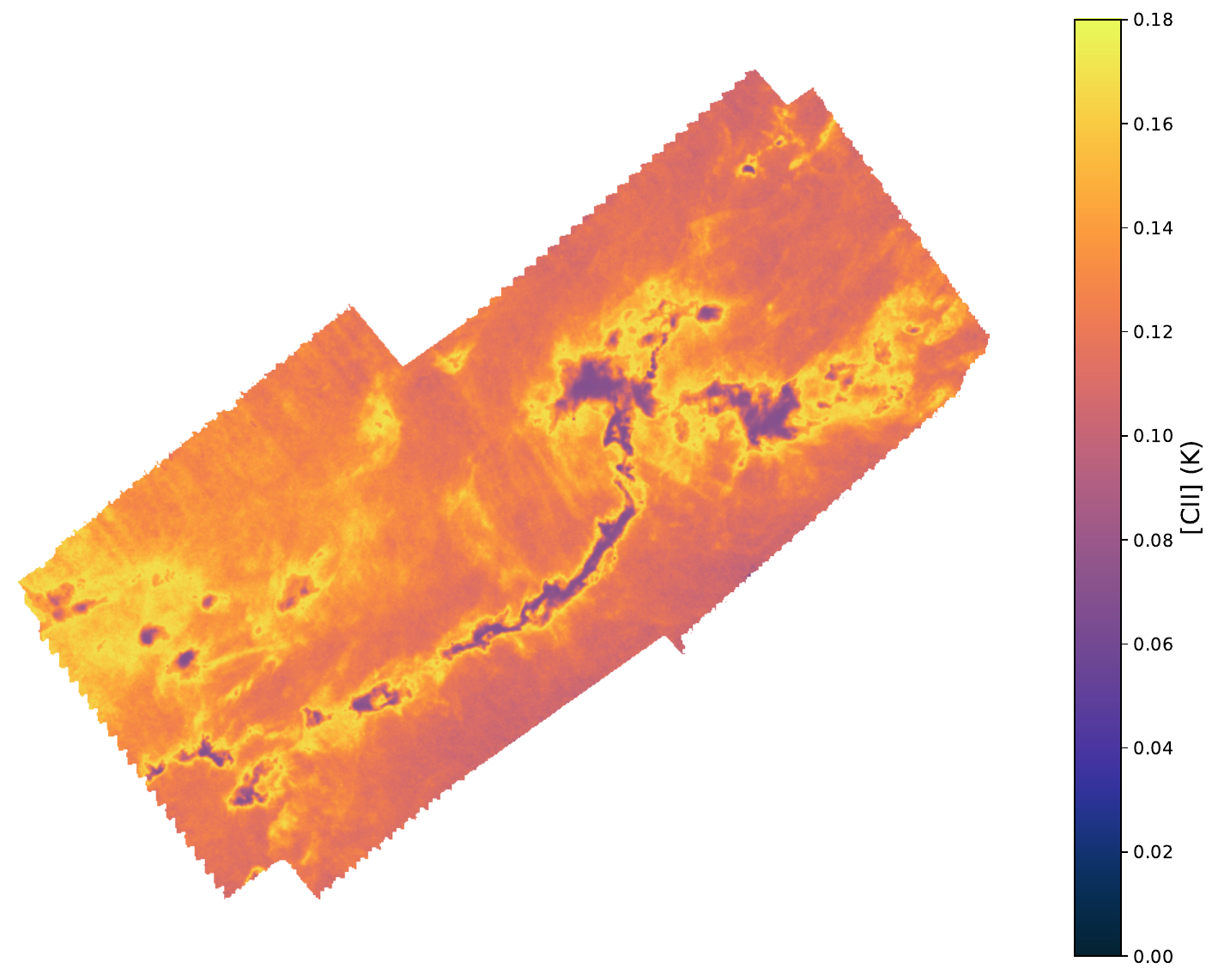}
    \includegraphics[width=0.32\textwidth]{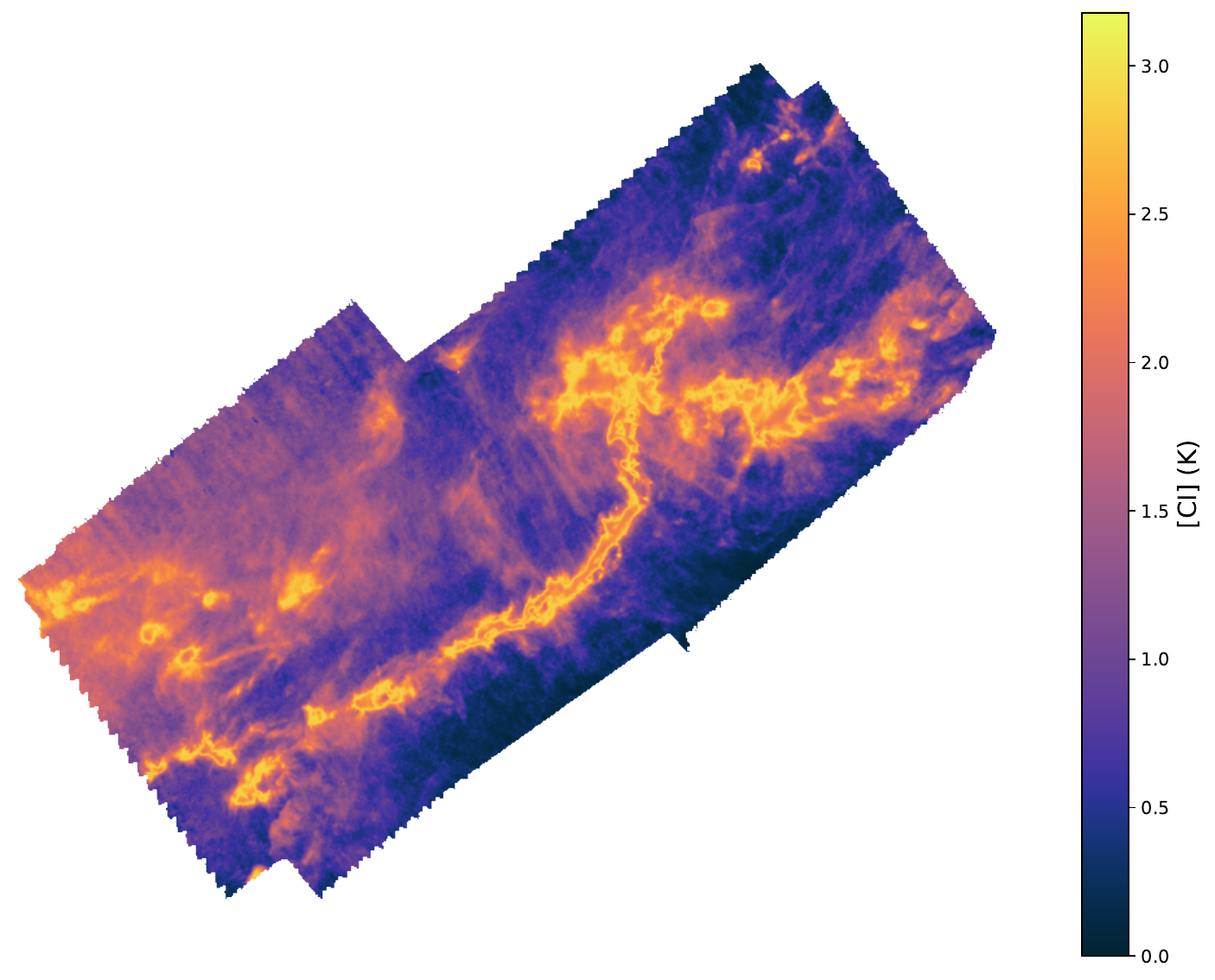}
    \includegraphics[width=0.32\textwidth]{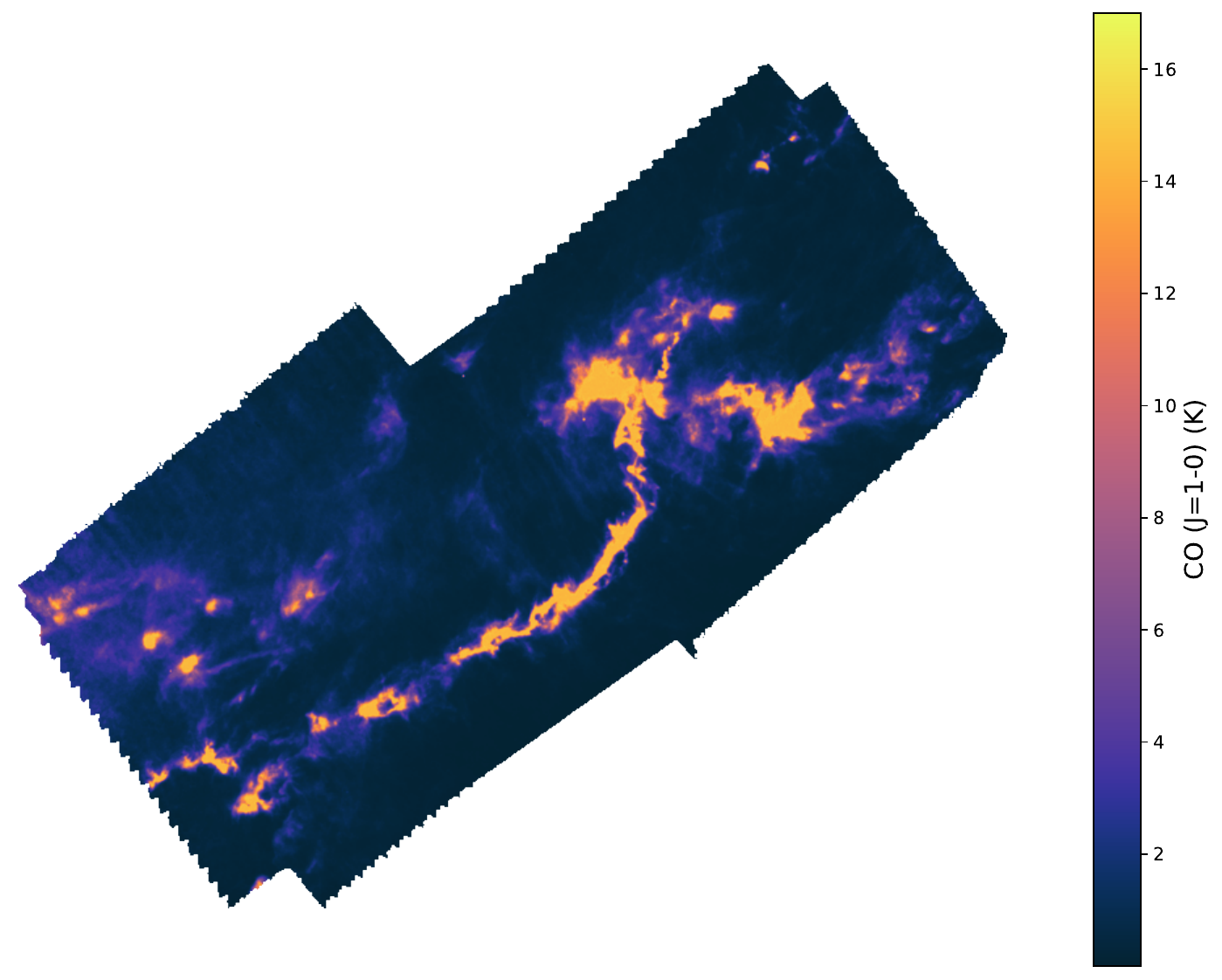}
    \caption{\label{fig:taurus_chem}Method I (top) and Method II (bottom) synthetic emission maps of the brightness temperature of [C II] 1900 GHz (left), [C I] 492 GHz, and CO (J = 1-0) 115 GHz emission.}
\end{figure*}

Figure \ref{fig:taurus_chem} shows the synthetic emission for [C II], [C I], and CO (J = 1-0) across the entire map for Method I (top) and Method II (bottom). To produce the final maps, the grid of brightness temperatures was interpolated for the pixel's values of ($n_J$, $N_{\rm eff}$). Further, since emission is expected even from diffuse regions, we added a component with an effective column density equivalent to the average turbulent column density interpolated for the pixel's values of ($n_T$, $<N_T>$). The emission shows the behavior broadly expected from one-dimensional PDR slabs, namely that the [C II] is dominant in the diffuse regions but greatly reduced toward the dense filament, [C I] is focused on the boundaries, while CO is primarily in the dense gas. The predicted Method II CO emission is more concentrated compared to the observed CO (J=1-0) map of \citet[][see Figure 5]{Narayanan2008}, but qualitatively matches the $^{13}$CO distribution. This may indicate that the predicted Method II $N_{\rm eff}$ in the diffuse regions is underestimated, or that the radiation temperature calculations are underpredicting the total line flux since this is not a velocity-resolved calculation of the line emission. 

\begin{figure}
    \centering
    \includegraphics[width=0.5\textwidth]{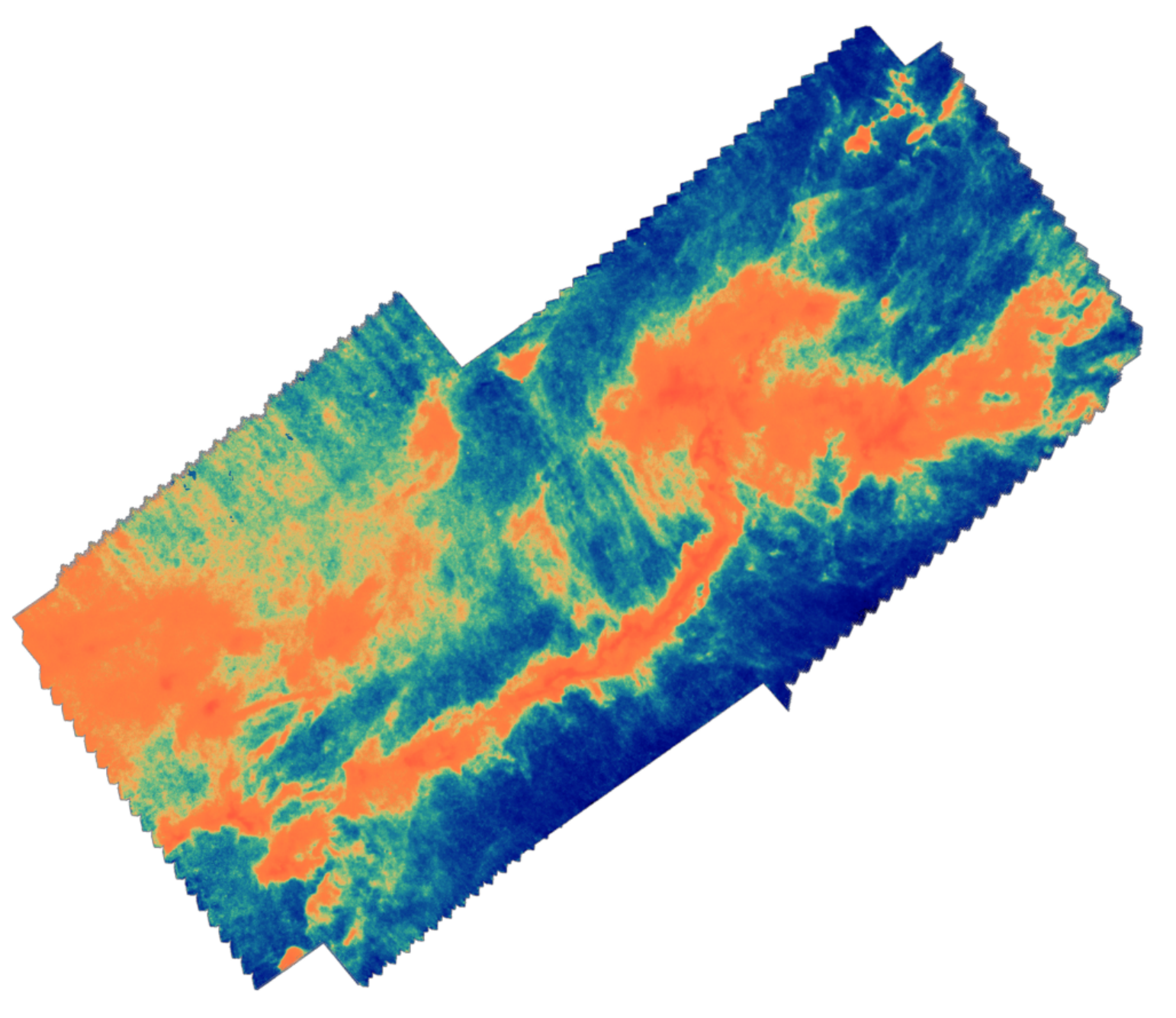}
    \includegraphics[width=0.5\textwidth]{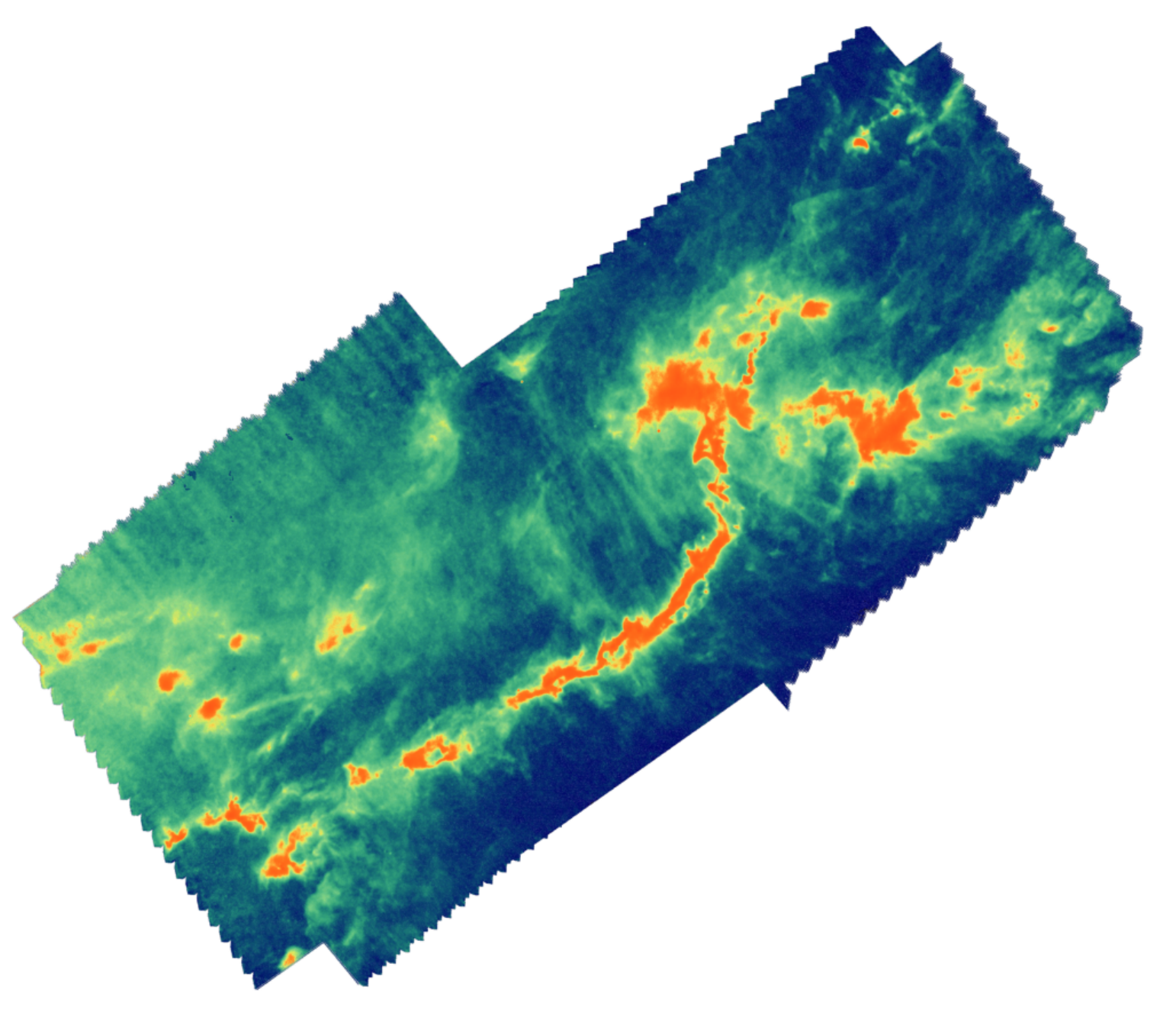}
    \caption{\label{fig:taurus_rgb}RGB image of Taurus with [C II] (blue), [C I] (green), and CO (J = 1-0) (red) for Method I (top) and Method II (bottom). The different lines have been given slight weights to highlight morphological differences.}
\end{figure}

Figure \ref{fig:taurus_rgb} shows a rendering of the synthetic emission maps as RGB images, with the same scalings used for each method, showing the different regions where the emission is dominating. The RGB images more concretely demonstrate the differences between the two methods. The lower effective column densities throughout the region using Region II lead to a reduction in CO, concentrating its emission tightly along the filament. The emission from [CI] and CO (J = 1-0) are more mixed in Method I and distinct in Method II. In comparison to the $^{12}$CO and $^{13}$CO emission from \citet{Narayanan2008}, Method I better reproduces the morphology of the emission, and therefore is more recommended for use with these types of astrochemical models. Such predictions could be of great interest for planning future observations, as long as regions have calculated total hydrogen column densities. On 24 threads, the production of the three maps took approximately 15 minutes, highlighting the utility of this approach in rapidly producing large predicted maps of emission.

\subsection{Assumptions and limitations important for use}
There are three key underlying assumptions of the model: i) gas temperature, ii) cloud thickness, and iii) the bulk properties of the turbulence. While these are key assumptions that will have an impact on the result, the speed of calculation and simplicity of the models make understanding their role and impact easier. In particular, the chosen transition density varies with the temperature, through $c_s^2$; the cloud depth as $L^-2$; and the sonic Mach number as $M_s^4$, which further depends on assumptions regarding the physics underlying the N-PDF. However, there is little to prevent a substitution for a different turbulent model of the N-PDF: it is a free choice of the end user. In fact, any distribution function could be used for the turbulent component, since it must be sampled with the prior that it must be less than the observed column density. For best performance and physical interpretability, the chosen distribution function should match as close as possible the expected turbulent component. While we utilized the Taurus fit from \citet{Burkhart2012} to estimate $M_s^4$, one could instead use the isothermal magneto-hydrodynamic turbulence ($A = 0.11$, $b = 1/3$) which was found to represent both super- and sub-Alfvenic turbulence simulations in \citet{Burkhart2012}. 

In the models presented here, we have chosen to use a fixed temperature of 15 K for all models. In reality, clouds are not isothermal. The temperature impacts both the estimation of $n_{\rm tr}$ linearly with the gas temperature and the gravitational column density as $N_J \propto T^{1/2}$. Therefore, increasing the gas temperature will physically inhibit fragmentation and gravitational collapse, and hence increase the transition density, therefore reducing the switch for intermediate densities, but also decreases the gravitational density through the gravitational column density. However, the description for the transition density would have to adapted away from isothermal turbulence, since the Mach number is kept constant throughout the cloud for the underlying model. For a given observed column density and sampled turbulent column density, $\Delta N \propto T^{1/2} n_H^{1/2}$ in the dense gas, and thus for a given $\Delta N$, an increase in $T$ will lead to a decrease in $n_H$. These two factors will partially work against each other, mitigating the impact, although the transition density scales linearly with temperature. Decreases in temperature will act in the opposite manner, as gas can more easily fragment and collapse gravitationally. There are different ways to alleviate the impact of this assumption. First, a density-dependent empirical temperature profile could be used. Second, the gas temperature could be assumed to equal the dust temperature and dust temperature maps constrained by observations could be used. Finally, the calculation could be run for different constant temperatures to get a range of possible densities. 

The calculations used in our model also require an estimation of the overall cloud depth. For some clouds, such as those in the study of \citet{Zucker2021}, this can be well estimated. For other regions, best-guess cloud depths must be used, as is done in our case for the Polaris Flare. There are two primary ways in which the cloud depth will impact the results. First, because the transition density is dependent on $L^{-2}$, assuming too large of a depth will decrease the transition density, and lead to more dense gas, since the criterion $n_H > n_{\rm tr}$ will be met at lower density, turning on the switch at lower densities. Second, the estimated densities are computed from the two column density components, with the turbulent cloud density scaling as $L^{-1}$. Therefore, a larger cloud depth will result in both a lower turbulent density and a higher gravitational density, and the opposite for a smaller cloud depth. However, the speed at which these models can be run allows for a range of cloud depths to be tested.

Finally, it is worth considering the limits of when this method is best used. The method requires a high-fidelity estimation of the hydrogen nuclei column density. Whatever errors or biases exist in such a column density map will propagate to the estimated number densities. The underlying method also assumes that each line of sight can be decomposed into a turbulent and gravitational component. This breaks down if there are many bound structures along the line of sight or if additional processes are becoming dominant such as the magnetic field and radiation pressure or feedback processes. We recommend the use of this method primarily for younger molecular clouds before the dynamics become dominated by feedback processes. 

The strength of the model, despite the underlying assumptions and limitations, is the speed of its use and the interpretability of the final estimated number densities. The simplicity of the model means that these biases can be better understood and that the decomposed number densities or column densities can be interpreted from a physics-forward viewpoint.

\section{Discussion and conclusions}\label{sec:discussion}
We presented here a novel method to estimate number density in GMCs from observed column density maps. The method is based on a simple physical model of decomposing the turbulent and gravitational components of the gas along the line of sight using assumptions of the properties of the turbulence within the molecular cloud. The method is applied to two molecular cloud regions, Taurus and the Polaris Flare, and compared against a simulation from the {\sc Starforge} suite. The method produces reasonable results for the turbulent and gravitational densities, with the former typically tracing the diffuse envelope and the latter the denser structures along each line of sight. Further, the model can predict an effective attenuating column density which can be used for astrochemical modeling. By comparing the predicted values with a simulated molecular cloud, we find that the turbulent density effectively traces the volume-weighted average density and the gravitational density is correlated with the mass-weighted average density. For detailed statistical modeling, Method III can be used to produce probability distributions of these two different density components and effective column densities which will enable a statistical modeling of molecular line and dust continuum emission for selected regions.

The underlying physical method is simple, requiring properties of the turbulence measured from the observed N-PDF and assumptions of the nature of the turbulence. The method thus ignores the impact of magnetic fields, which can play an important role in the dynamics of molecular clouds (see for instance \citet{Skalidis2023} for the Polaris Flare). However, one could approximate these effects by changing the model for the turbulence that is assumed in the estimation of the turbulent column density. Further, the method requires an estimation of cloud thicknesses. However, with the launch of Gaia, this is becoming more feasible with robust three-dimensional cloud maps \citep[][]{Leike2019, Leike2020,  Zucker2021, Dharmawardena2022, RezaeiKh2022, Edenhofer2024, RezaeiKh2024}. Finally, in our initial estimations, we assumed the gas in the cloud was isothermal. While this assumption makes the estimations more straightforward, the method is general enough that along each line of sight one could use an estimated gas temperature, such as by assuming the gas and dust temperatures are equal, or using gas temperatures derived using atomic and molecular line emission. A final constraint is the ability to characterize the bulk turbulence properties of the cloud via fitting a log-normal N-PDF to the observed column densities, although the computed N-PDF has potential biases due to lack of completeness at low column densities, foreground and background gas, and noise \citep{Alves2017, Kortgen2019}.

The turbulent and gravitational densities we predicted are in qualitative agreement both with the properties of the Polaris Flare and Taurus. Further, we find a peak gas density in the Taurus L1495/B213 filament of $n_J \approx 10^5$ cm$^{-3}$, which is in agreement with previous estimations \citep{Li2012, Xu2023} when accounting for the difference in resolutions used in the studies. We find that for the Polaris Flare, when using the Method II turbulent-density density, the gravitational density primarily peaks within the saxophone feature, which is in agreement with the idea that this feature is the main dense-gas component \citep{Skalidis2023}, with much of the rest of the density being enhanced in the turbulent component. 

In conclusion, we have presented a novel method for estimating the hydrogen-nuclei number density of GMCs by assuming the gas along the line of sight can be decomposed into a turbulence-dominated and gravitationally dominated component. By utilizing the method in reverse, from the dense-gas estimations we can further estimate effective attenuating column densities. Thus, from an observed column density map and assumptions regarding the nature of the turbulence, we can estimate the decomposition of the gas along the line of sight into different components and the effective column density at the native resolution of the observations. The method presented is general, such that extensions for additional physics or relaxing assumptions can be readily done. Importantly, our model enables the use of data at the native resolution without the need to downsample, and the methods are intrinsically parallel. The method is also embarrassingly parallel, allowing estimations across entire cloud regions to be rapid enough to do parameter explorations. We believe this method can be a useful and flexible tool for the star formation and astrochemistry communities in estimating the number density for understanding the dynamics and chemical evolution of GMCs. 

\section{Data availability}
We have created an example Jupyter notebook in a public GitHub repository hosted at \url{https://github.com/AstroBrandt/GMCDensityEstimation}, archived as DOI:10.5281/zenodo.14754248. using a low-resolution version of the Taurus map which can be easily run on a laptop so potential users of the model can explore the various parameters and methods. 

\begin{acknowledgements}
We thank Jouni Kainulainen and Gina Panopoulou for interesting and thoughtful discussions which improved this work and Thomas Bisbas for providing the PDR chemical model grids (DOI:10.5281/zenodo.7310833). BALG is supported by the Chalmers Initiative on Cosmic Origins as Cosmic Origins Postdoctoral Fellows. Support for MYG was provided by NASA through the NASA Hubble Fellowship grant \#HST-HF2-51479 awarded by the  Space  Telescope  Science  Institute,  which is operated by the Association of Universities for Research in Astronomy, Inc.,  for NASA, under contract NAS5-26555. The simulation analyzed was run on the Frontera supercomputer at the Texas Advanced Computing Center through award AST21002. This research is part of the Frontera computing project at the Texas Advanced Computing Center. Frontera is made possible by the National Science Foundation award OAC-1818253.
\end{acknowledgements}

\bibliographystyle{aa}
\bibliography{lib} 

\onecolumn
\begin{appendix}
\section{Additional figures}
This appendix contains the figures presenting the Model I and Model II results for the Polaris Flare. 

\begin{figure*}[htb!]
    \centering
    \includegraphics[width=0.48\textwidth]{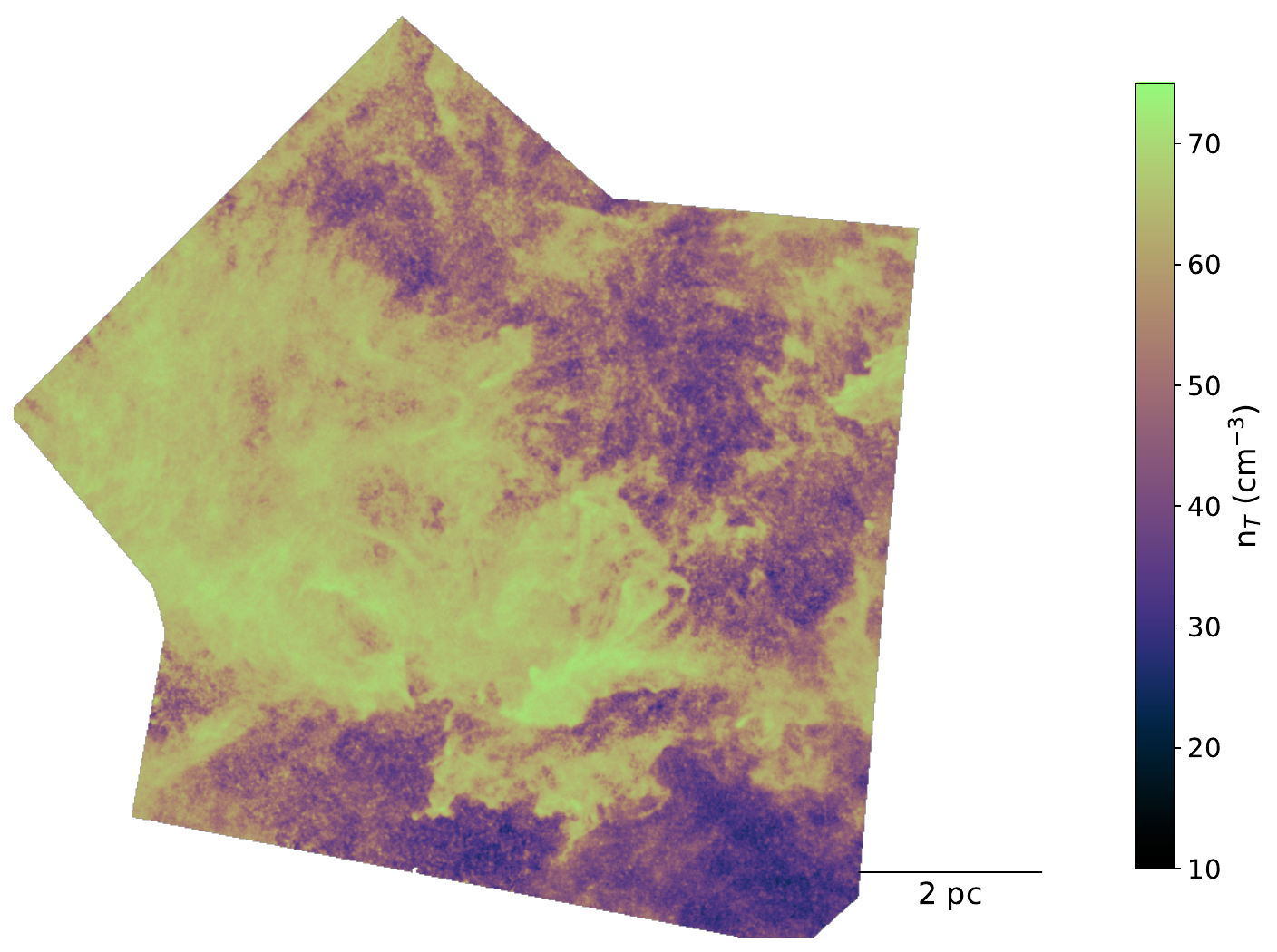}
    \includegraphics[width=0.48\textwidth]{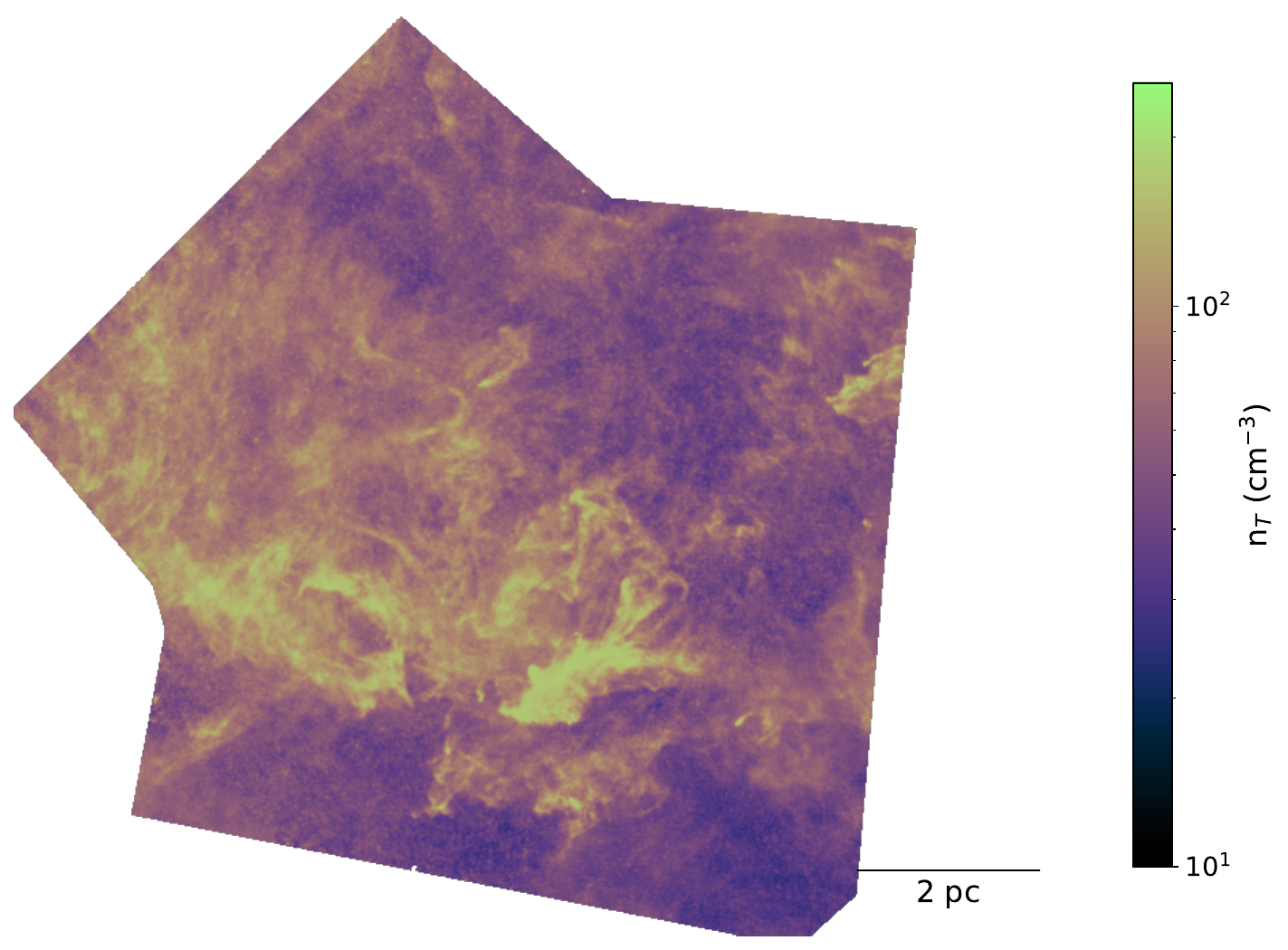}
    \caption{\label{fig:polaris_turb} Same as Figure \ref{fig:taurus_turb} but for Polaris.}
\end{figure*}

\begin{figure*}[htb!]
    \centering
    \includegraphics[width=0.48\textwidth]{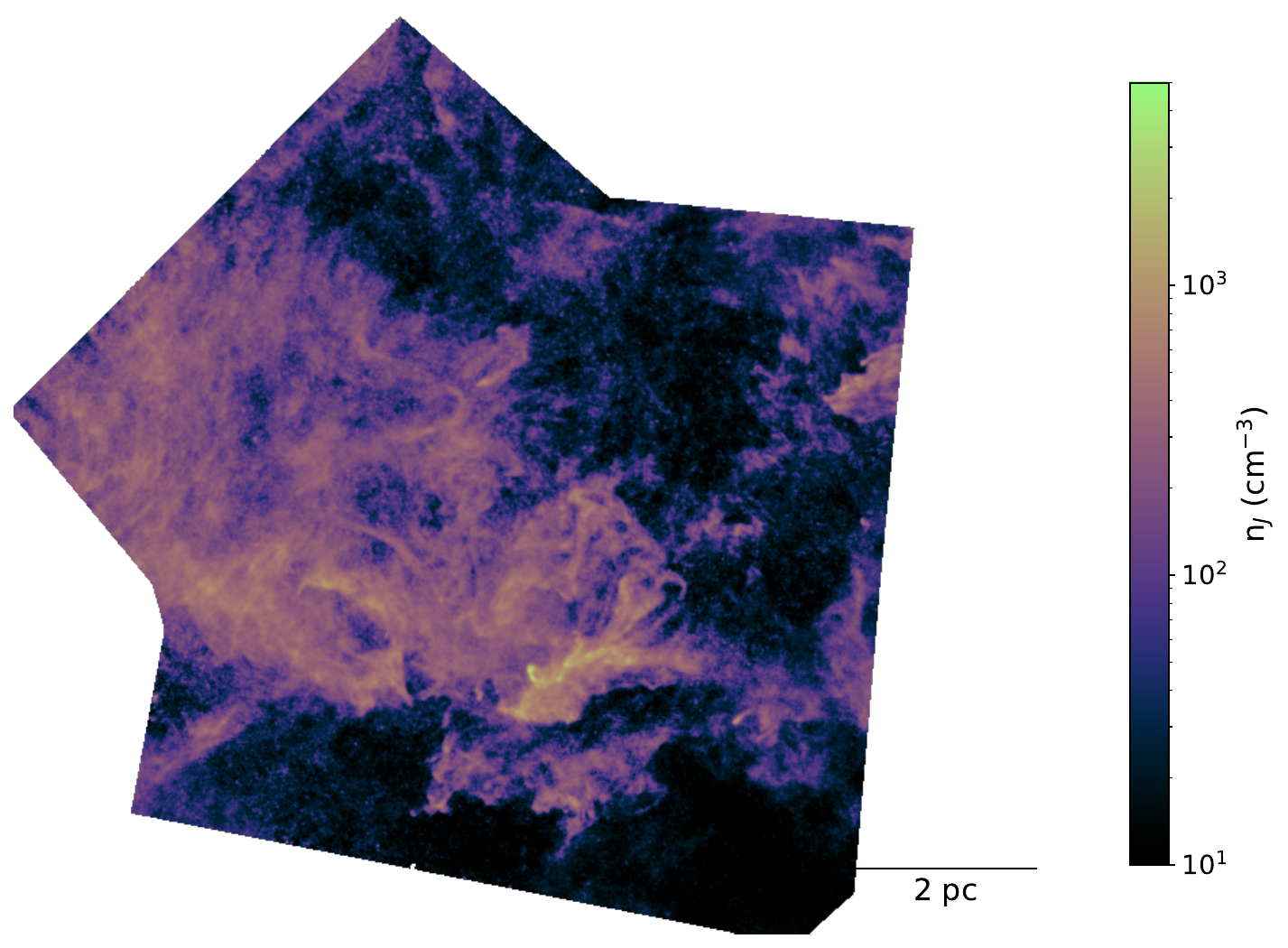}
    \includegraphics[width=0.48\textwidth]{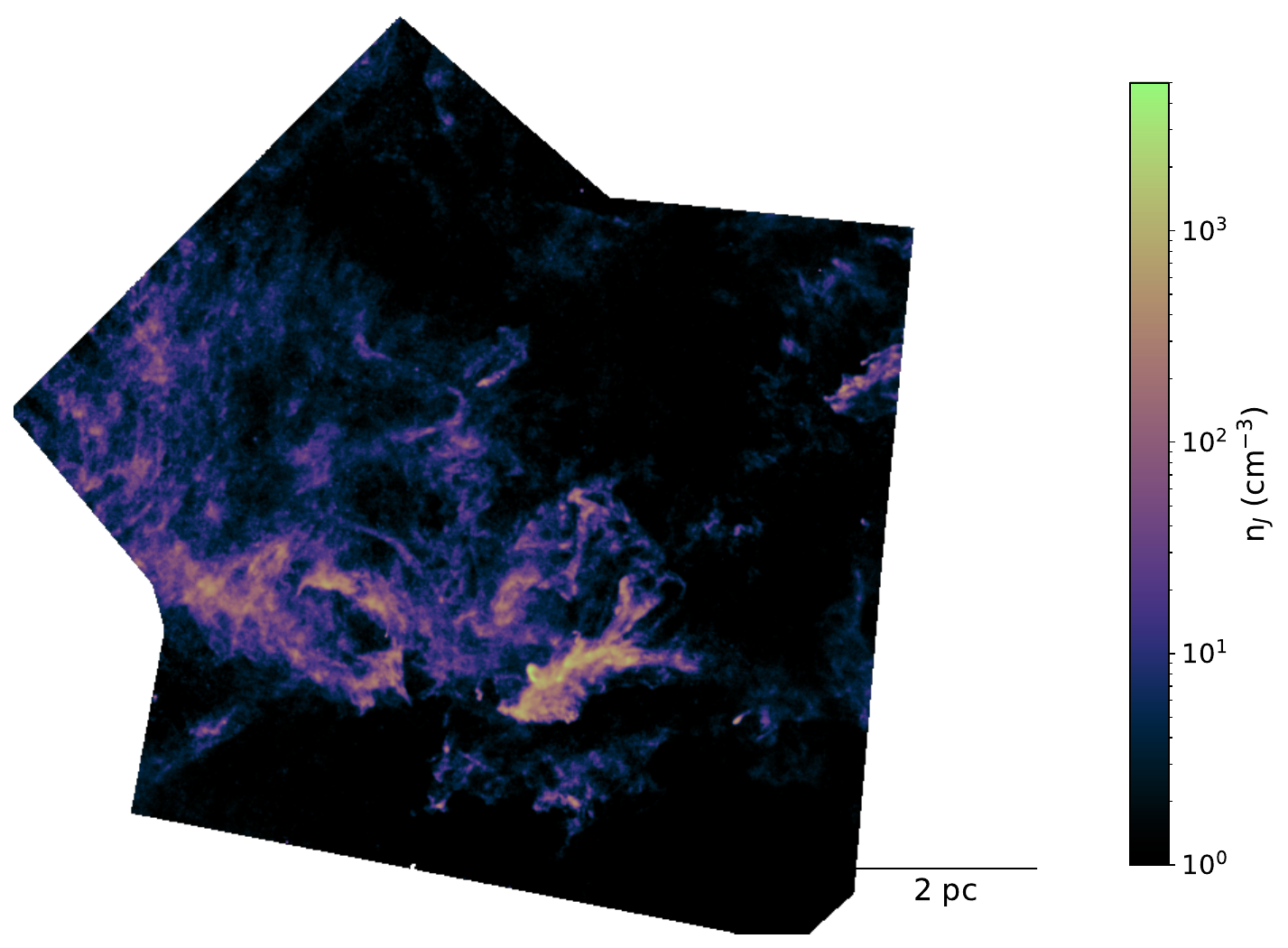}
    \caption{\label{fig:polaris_grav} Same as Figure \ref{fig:taurus_grav} but for Polaris.}
\end{figure*}

\begin{figure*}[htb!]
    \centering
    \includegraphics[width=0.48\textwidth]{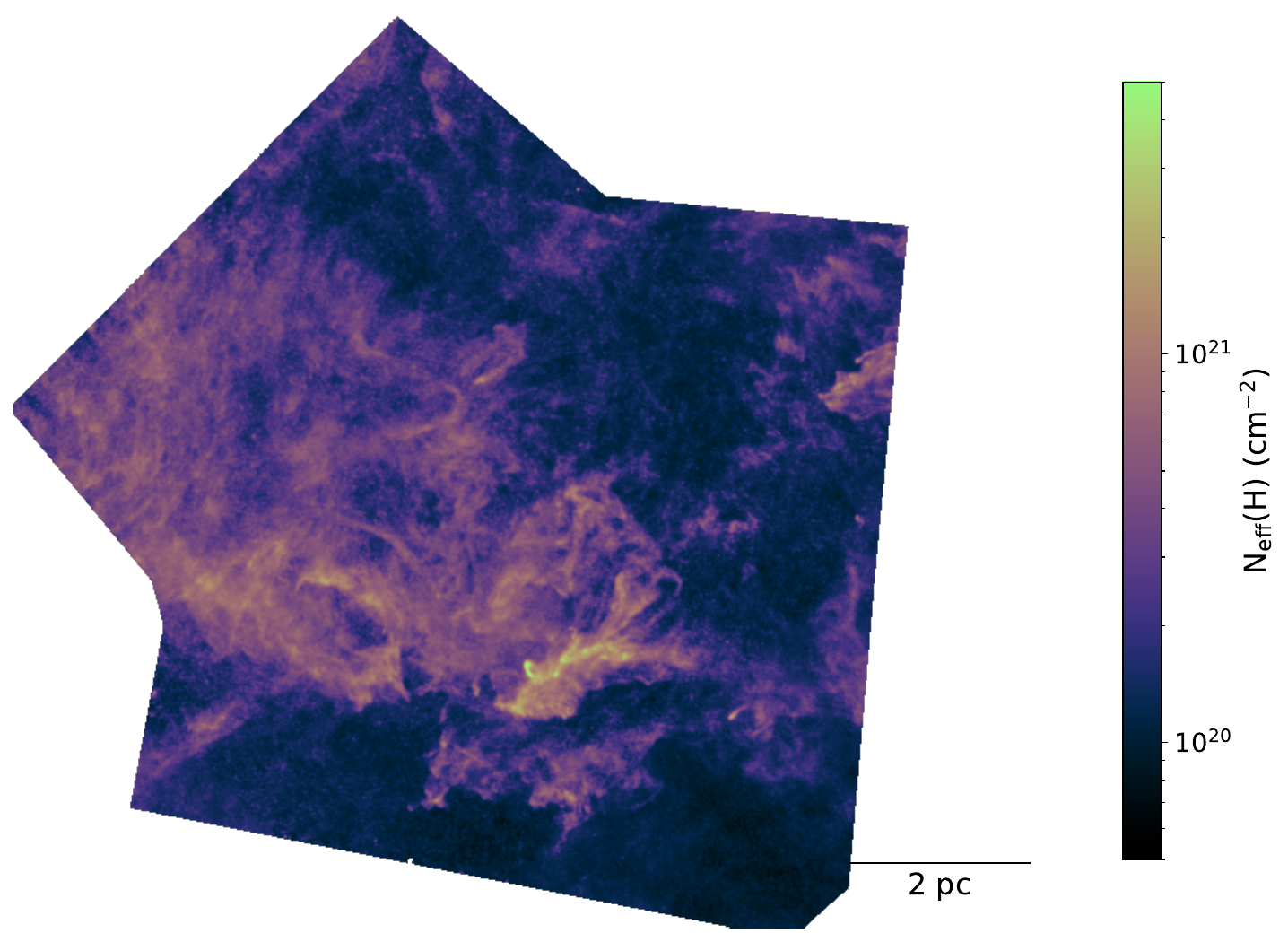}
    \includegraphics[width=0.48\textwidth]{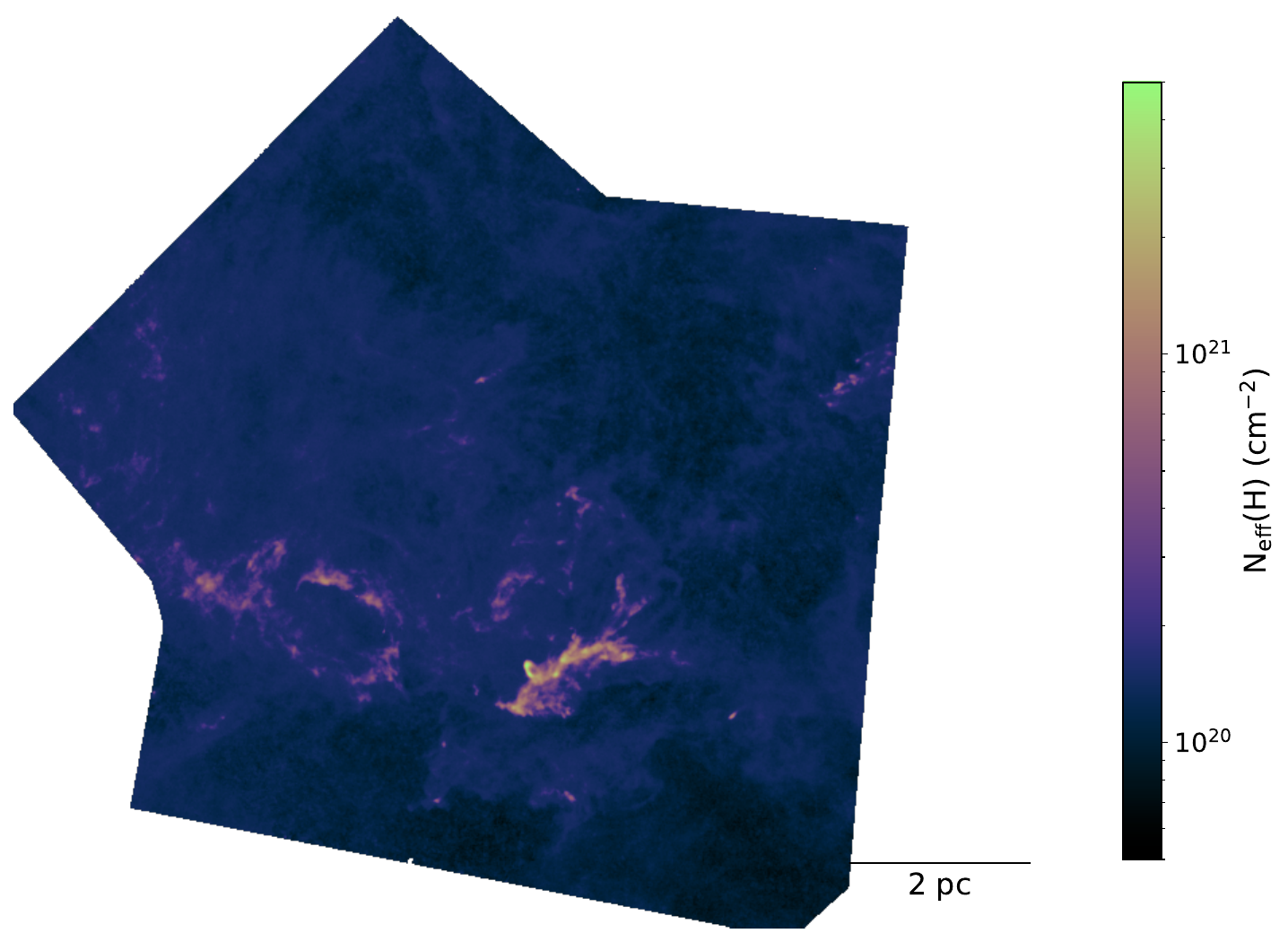}
    \caption{\label{fig:polaris_neff} Same as Figure \ref{fig:taurus_neff} but for Polaris.}
\end{figure*}

\begin{figure*}[htb!]
    \centering
    \includegraphics[width=0.48\textwidth]{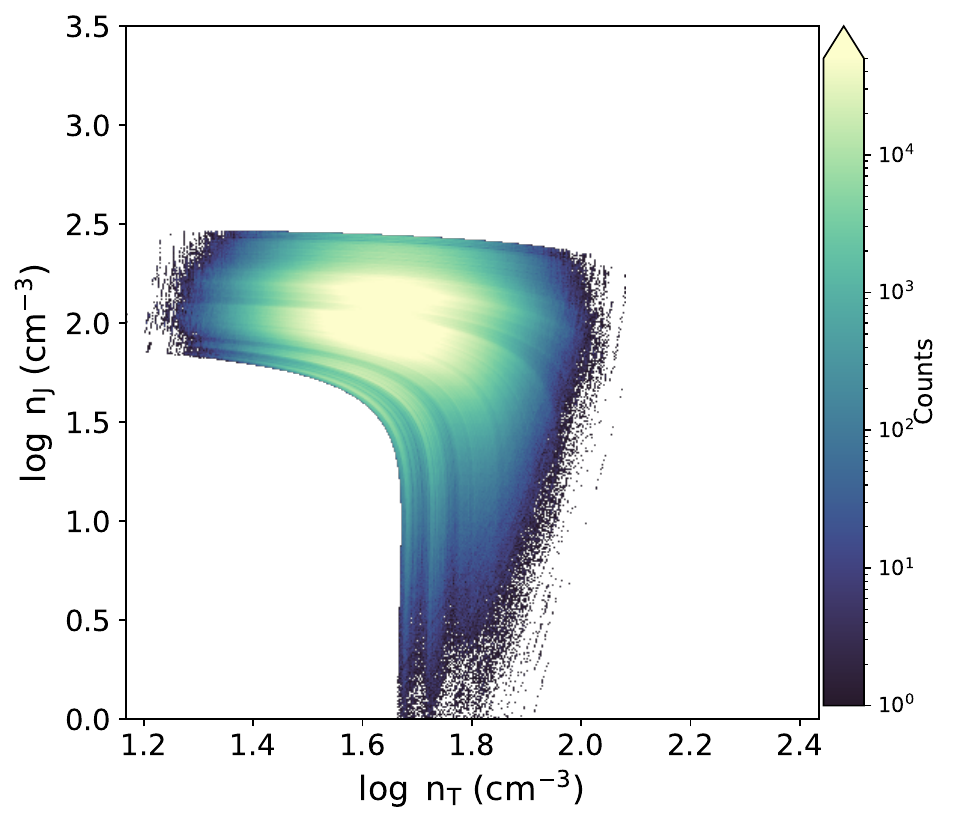}
    \includegraphics[width=0.48\textwidth]{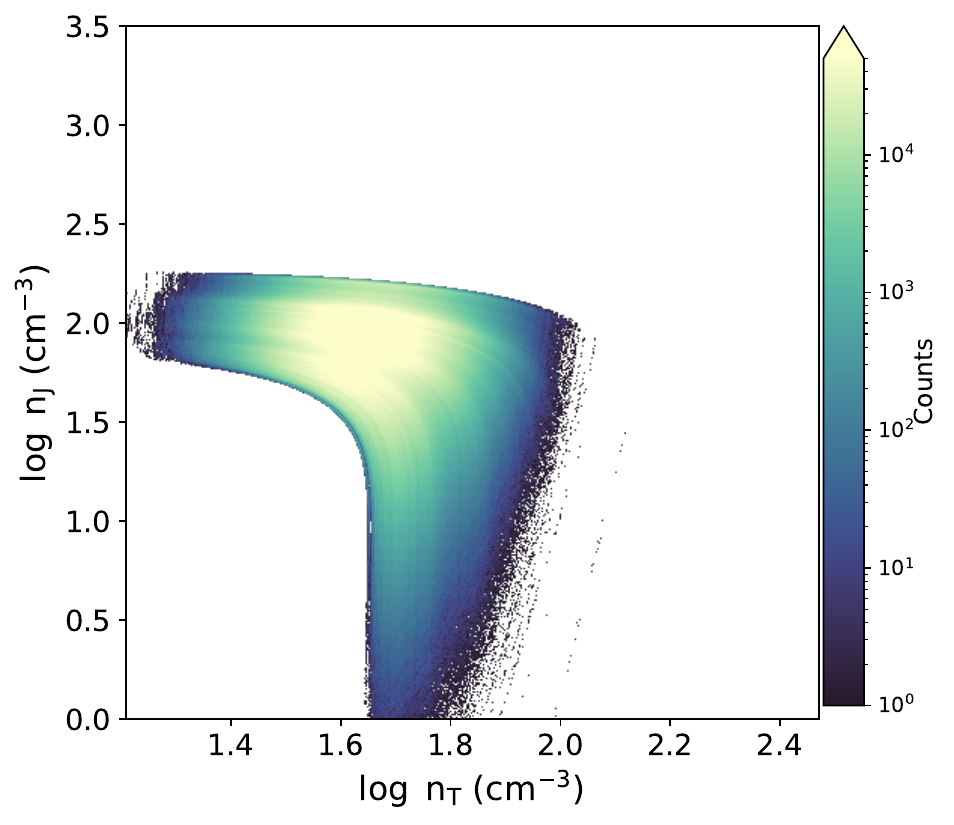}
    \caption{\label{fig:polaris_bivariate} Same as Figure \ref{fig:taurus_bivariate} but for Polaris.}
\end{figure*}

\end{appendix}

\end{document}